\documentclass[twocolumn]{revtex4-1}   	        
\usepackage{geometry}                		
\usepackage{graphicx}				
\usepackage{epsfig}				
\newcommand{\beq}{\begin{eqnarray}}
\newcommand{\eeq}{\end{eqnarray}}
\begin{document}

\title{ Analytical Theory of Neutrino Oscillations in Matter with CP violation }
\author{Mikkel B. Johnson \\
Los Alamos National Laboratory, Los Alamos, NM 87545 \\
Ernest M. Henley\\
Department of Physics, University of Washington, Seattle, WA 98195\\ 
Leonard S. Kisslinger\\
Department of Physics, Carnegie-Mellon University, Pittsburgh, PA 15213\\}

\begin{abstract}

We develop an exact analytical  formulation of neutrino oscillations in matter within the framework of the Standard Neutrino Model assuming 3 Dirac Neutrinos.
Our Hamiltonian formulation, which includes CP violation, leads to expressions for the partial oscillation probabilities that are linear combinations of spherical Bessel functions in the eigenvalue differences.  The coefficients of these Bessel functions are polynomials in the neutrino CKM matrix elements, the neutrino mass differences squared, the strength of the neutrino interaction with matter, and the neutrino mass eigenvalues in matter. We give   exact  closed-form expressions for all partial oscillation probabilities in terms of these basic quantities.
Adopting the Standard Neutrino Model, we then examine how the exact expressions for the partial oscillation probabilities might simplify by expanding in one of the small parameters $\alpha$ and $\sin\theta_{13}$ of this model.  We show explicitly that for small $\alpha$ and $\sin\theta_{13}$ there are branch points in the analytic structure of the eigenvalues that lead to singular behavior of expansions near the solar and atmospheric resonances. We present numerical calculations that indicate how to use the small-parameter expansions in practice.

\end{abstract}

\maketitle

\noindent

PACS Indices: 11.3.Er,14.60.Lm,13.15.+g

\vspace{1mm}

\noindent

Keywords: 

\vspace{1mm}

\noindent

\section{Introduction}

In this paper, we develop an exact analytical representation of neutrino oscillations~\cite{wolf} in matter within the framework of the Standard Neutrino Model (SNM)~\cite{ISS}  with 3 Dirac Neutrinos.  The exact closed-form expressions we give for the time-evolution operator $S(t,t')$ are obtained from $H_\nu$  using the Lagrange interpolation formula given in Ref.~\cite{barg}.  The resulting expressions are easily evaluated without any approximations.

The paper is divided into two main parts. In the first, we summarize the main results of our theory. Details underlying the derivation are given in Appendices. We also retrieve the well-known two-neutrino flavor results as a special case of our general results.

In the second part we address other analytical formulations found in the literature.  The expansion of the neutrino oscillation probability in one of the small parameters $\alpha$ and $\sin^2\theta_{13}$ of the standard neutrino model (SNM) for $H_\nu$ is of particular interest. The seminal work along these lines is found in Refs.~\cite{f,ahlo,JO}.  This work  underlies many of the analyses and exploratory studies of experiments at present and future neutrino facilities,  including our earlier work~\cite{hjk1,khj2,hjk1a,khj1E}.  

The present paper was undertaken, and used, for the purpose of independently confirming the results of Refs.~\cite{hjk1,khj2,hjk1a,khj1E}.  
We find that  the accuracy of the expanded oscillation probability is restricted by the presence of branch points in the analytic structure of the eigenvalues of neutrinos propagating in matter.  We also show that the regions where the expanded results are reliable is different for expansions in $\alpha$~\cite{f} and $\sin^2\theta_{13}$~\cite{ahlo,JO}. We then map out regions where the expanded results  are reliable by comparing numerical results to the exact results of our Hamiltonian formulation.  

Another recent study~\cite{kiss1} takes a complementary approach and finds that the predictions of Refs.~\cite{hjk1,khj2,hjk1a,khj1E} can be improved in certain regions using an exact evaluation of the integral $I_{\alpha *}$ rather than the approximate one found there.  It concludes that within these regions, predictions of  $(\mu,~e)$ oscillations improve for certain values of the  experimental parameters.

The dimensionalities of the neutrino Hamiltonian $H_\nu$ and the parameter space characterizing the mixing of three neutrino pairs are  sources of difficulty for  finding a tractable representation of the oscillation probability.  The Lagrange interpolation formula~\cite{barg} is enormously helpful, providing an exact and  formally elegant expression for the exponentiation of an $n\times n$  matrix. 

The description of  two-flavor neutrino oscillations is elementary by comparison.  In that case,   $H_\nu$  is a $2\times 2$ matrix, and the mixing is described by a single real parameter. 

\section { Neutrino Dynamics }
\label{ND}

We will be interested in the dynamics of the three known neutrinos and their corresponding anti-neutrinos in matter. This dynamics is determined by the time-dependent Schr$\ddot{o}$dinger equation,
\begin{eqnarray}
\label{tdse}
i\frac{d}{dt}|\nu(t)> &=& H_\nu |\nu(t)> ~,
\end{eqnarray}
where the neutrino Hamiltonian, 
\beq
\label{fulh}
 H_\nu=H_{0v}+H_{1} ~,
\eeq
consists of a piece $H_{0v}$ describing neutrinos in the vacuum and a piece $H_{1}$ describing their interaction with matter.

The solutions of Eq.~(\ref{tdse})  may be expressed in terms of stationary-state solutions of the eigenvalue (EV) equation
\beq
\label{eveq}
H_\nu|\nu_{mi}> &=& E_i |\nu_{mi}> ~,
\eeq
where the label ``$m$" indicates neutrino mass eigenstates, as distinguished from their flavor states sometimes denoted the label ``$f$". 
In operator form, this dynamics may be expressed in terms of the time-evolution operator $S (t',t)$, which describes completely the evolution of states from time $t$ to $t'$ and also satisfies the time-dependent Schr$\ddot{o}$dinger equation.

We will examine neutrinos propagating in a uniform medium for interactions constant not only in space but also time. Because the Hamiltonian is then translationally invariant, attention may be restricted to states, both in the vacuum and in matter, characterized by momentum $\vec p$ and therefore having the overall $r$-dependence $e^{i\vec p\cdot\vec r}$. In this case expressions may be simplified by suppressing the overall plane wave, a convention we adopt. 

For time-independent interactions, $S(t',t)$,
\beq
\label{smtx}
S(t',t) &=& e^{-iH_\nu (t' - t)} ~,
\eeq
depends on time  only through the time {\it difference} $t'-t$. Then, written in terms of the stationary state solutions $|\nu_{mi}>$ of Eq.~(\ref{tdse}), 
\beq
\label{tunif}
S(t',t) &=& \sum_i |\nu_{mi}> e^{-iE_i (t' - t ) }<\nu_{mi}| ~.
\eeq

With the momentum dependence factored out, three basis states $|M(i)>,~i=(1,2,3)$ are then required to describe three neutrinos. The basis states  correspond to a specific representation, as in descriptions of a spin-1 object.  The basis should, of course, be orthogonal, 
\beq
<M(i)|M(j)> &=& \delta_{ij} 
\eeq
and complete,
\beq
\sum_{k} |M(k)><M(k)| = \bf 1 ~.
\eeq

Once the basis is chosen, wavefunctions for a neutrino state are naturally introduced as the components of this state in the chosen basis. For example, with the eigenstates of Eq.~(\ref{eveq}) expanded in the basis,
\beq
\label{nudef}
|\nu_{mj}> &=&  \sum_i |M(i)> m^{i}_{j} ~,
\eeq
the wave functions of $|\nu_{mj}>$ would be the set $m^{i}_{j},~i = (1,2,3)$. With the plane wave factored out, the wave function is just a set of three numbers. 
Additionally, introduction of a basis makes it possible to represent neutrino states and operators such as $H_\nu$  in matrix form, with each entry in the matrix corresponding to a projection of the object being described onto the basis.

In this paper we take the Hamiltonian in Eq.~(\ref{fulh}) to be  expressed in the {\it standard representation}, where the mass basis states $|M(i)>$ are taken as the set of states  that diagonalize the neutrino vacuum Hamiltonian $H_{0v}$, {\it i.e.} $|M(i)> \equiv |\nu^0_{mi}> = |{\bar \nu}^0_{mi}>$,  
\beq
\label{MBS}
H_{0v}|\nu^0_{mi}> = E^0_i |\nu^0_{mi}> ~.
\eeq
In matrix form 
\beq
\label{h0st}
H_{0v} &=& \left( \begin{array}{ccc} 
E^0_3 & 0 & 0 \\ 0 & E^0_2
& 0 \\ 0 & 0 & E^0_1 \end{array} \right )  ~,
\eeq
with the EV's taken to be ordered $E^0_1 \leq E^0_2 \leq   E^0_3$ as in the normal mass hierarchy. In the literature, the Hamiltonian is often expressed in a different basis obtained by rotating to one  in which the complete neutrino Hamiltonian is diagonal as in Ref.~\cite{f}.  

We assume here that that neutrinos and anti-neutrinos represented by $|\nu^0_{mi}>$ and $|{\bar \nu}^0_{mi}>$, respectively, are the structureless elementary Dirac fields of the the Standard Neutrino Model~\cite{ISS}.  For this reason the theory is invariant under CPT, so the mass of an anti-neutrino in the vacuum is the same as that for its corresponding neutrino.

\subsection {Flavor and Mass States}

Neutrinos are produced and detected in states of good flavor, $|\nu_{fi}>$.  The three flavors, electron ($e$), muon ($\mu$), and tau ($\tau$) correspond, respectively, to the index values $i=(1,2,3)$.  In the vacuum, each flavor state is a specific linear combination of the three mass eigenstates $|M(i)>$ of the neutrino vacuum Hamiltonian $H_{0v}$. This linear combination is expressed in terms of the same set of numbers $U_{ij}$ for both neutrinos and anti neutrinos
\beq
\label{Fdef}
|\nu^0_{fi}> &=& \sum_j U^*_{ij} |M(j)>  \nonumber \\
|{\bar \nu}^0_{fi}> &=& \sum_j U_{ij} |M(j)>  
~,
\eeq
where $U_{ij}$ are the elements of a unitary operator $U$, the neutrino analog of the familiar CKM matrix. It is standard to express $U_{ij}$ in terms of three mixing angles $(\theta_{12},\theta_{13},\theta_{23})$ and a phase $\delta_{cp}$ characterizing $CP$ violation,
\beq
\label{Ubase}
&\left( \begin{array}{ccc} c_{12} c_{13} & s_{12} c_{13} & s_{13} e^{-i\delta_{cp}} \\ U_{21} & U_{22} & s_{23} c_{13} \\ U_{31}  & U_{32} & c_{23} c_{13}  \end{array} \right ) ~,
\eeq 
where 
\beq
\label{Udef}
U_{21}&=& -s_{12} c_{23} - c_{12} s_{23} s_{13} e^{i\delta_{cp}} \nonumber \\
U_{22}&=& c_{12} c_{23} - s_{12} s_{23} s_{13} e^{i\delta_{cp}} \nonumber \\
U_{31}&=& s_{12} s_{23} - c_{12} c_{23} s_{13} e^{i\delta_{cp}} \nonumber \\
U_{32}&=& - c_{12} s_{23} - s_{12} c_{23} s_{13} e^{i\delta_{cp}}  ~.
\eeq
We use here the standard abbreviation $s_{12}\equiv\sin{\theta_{12}}$, $c_{12}\equiv\cos{\theta_{12}}$, {\it etc}. The parameters $\theta$ and $\delta_{cp}$ are determined from experiment. 

Because  $U_{ij}\rightarrow U^*_{ij}$ with $\delta_{cp}\rightarrow -\delta_{cp}$ it follows that the relationship in Eq.~(\ref{Fdef}) between flavor and mass states for anti-neutrinos and neutrinos in the vacuum is equivalent to $\delta_{cp} \leftrightarrow -\delta_{cp}$.

\subsection{Neutrino Interacting Hamiltonian}

The interaction $H_1$, determined by taking the electron flavor states scattering from the electrons of the medium to mediate the interaction, is then expressed as an operator in the standard representation, 
\beq
\label{vdef}
H_{1} &=& U^{-1} \left( \begin{array}{ccc} V & 0 & 0 \\ 0 & 0  & 0 \\ 0 & 0 & 0   \end{array} \right ) U ~,
\eeq
with $V = \pm\sqrt{2} G_F n_e $ and $n_e$ the electron number density in matter.

For electrically neutral matter consisting of protons, neutrons, and electrons, the electron density $n_e$ is the same as the proton density $n_p$, 
\beq
n_e &=& n_p \nonumber \\
&=& R N ~,
\eeq
where $N= n_n + n_p$ is the average total nucleon number density and $R= n_p/N$ is the average proton-nucleon ratio. In the earth's mantle, the dominant constituents of matter are the light elements so $R \approx 1/2$; in the surface of a neutron star $R<<1$.  Matrix elements of $H_1$ are thus
\beq
\label{mxh1h0}
<M(k)|H_{1}|M(k')> &=& U^*_{1k} V U_{1k'} ~.
\eeq

Matrix elements of $H_1$ are thus
\beq
\label{mxh1h0}
<M(k)|H_{1}|M(k')> &=& U^*_{1k} V U_{1k'} ~.
\eeq

\subsection{Dimensionless variables}

The results are most naturally expressed in dimensionless variables. We first take  advantage of the global phase invariance to express all energies relative to the vacuum EV $E^0_1$ of the same momentum. We indicate that a quantity is expressed relative to $E^0_1$ by placing a bar over it, {\it e.g.},
\beq
\bar E^0_i &\equiv& E^0_i - E^0_1 ~.
\eeq
We follow the same convention for the Hamiltonian,
\beq
{\bar H}_\nu &\equiv& H_\nu - {\bf 1} E^0_1 ~,
\eeq
so the EV equation Eq.~(\ref{eveq}) becomes
\beq
\label{eveq1}
 ( \bar H_{0v}  + H_1) |\nu_{mi}> &=& \bar E_i |\nu_{mi}>,
\eeq
where
\beq 
\label{h0st2}
\bar H_{0v} &\equiv& H_{0v} -{\bf 1} E^0_1   ~.
\eeq

Then, to express the theory in dimensionless variables we divide all energies, including the Hamiltonian, by $\bar E^0_{3} = E^0_3 - E^0_1$. The stationary-states $|\nu_{mi}>$ are also be determined from the dimensionless Hamiltonian ${\hat {\bar  H}}_\nu $,
\beq
\label{hamhat1}
{\hat {\bar  H}}_\nu &=& {\hat {\bar H}}_{0v} + {\hat H}_1 ~, 
\eeq 
{\it i.e.}, from the solutions of 
\beq
\label{eveq2}
{\hat {\bar  H}}_\nu |\nu_{mi}> &=& {\hat {\bar E}}_i |\nu_{mi}> ~,
\eeq
where the ``hat" placed over a quantity indicates it is dimensionless. 
Thus
\beq
\label{hatbarE}
{\hat {\bar E}}_i &\equiv& \frac{ \bar E_i   }{ \bar E^0_{3} } \nonumber \\ 
{\hat {\bar H}}_{0v} &\equiv& \frac{ {\bar H}_{0v} } { \bar E^0_{3} } ~,
\eeq
and ${\hat H}_1 $ is obtained from $H_1$ by replacing
\beq
\label{hatAdef}
V \rightarrow \hat A &\equiv& \frac{V }{\bar E^0_3} ~.
\eeq
The quantity ${\hat A}$ is the same as that defined in Refs.~\cite{f,hjk1,khj2,hjk1a,khj1E,kiss1}.
The connection of the Hamiltonian ${\hat {\bar  H}}_\nu $ to the full Hamiltonian $ H_\nu = H_{0v}+H_{1} $ is then
\beq
\label{fulh1}
 H_\nu &=& {\bf 1} E^0_1 + \bar E^0_{3} {\hat {\bar H}}_\nu ~.
\eeq

\subsection { Neutrino Vacuum Hamiltonian ${\hat {\bar H}}_{0v}$  }

The case of main interest for many situations  is  
the ultra-relativistic limit, $\vec |p| >> m^2$ (we take the speed of light $c=1$).  For ultra-relativistic neutrinos in the laboratory frame, the energy of a neutrino in the vacuum becomes
\beq 
\label{Evac1}
E^0_i &\approx&  |\vec p| + \frac{m_i^2}{2E} ~,
\eeq
where $m_i$ is its mass the vacuum.  Similarly, $E_i$ appearing in Eq.~(\ref{eveq}) may be written
\beq 
E_i &\approx&  |\vec p| + \frac{M_i^2}{2E} ~,
\eeq
where $M_i$ is its mass in the medium.    
Thus, in this limit,
\beq
{\hat {\bar E}}_i &\to& \frac{M_i^2 - m^{2}_1}{m^{2}_3 - m^{2}_1}
\eeq
and 
\beq 
\label{h0st2}
{\hat {\bar H}}_{0v} &\to & \left( \begin{array}{ccc} 0 & 0 & 0 \\ 0 & \alpha
& 0 \\ 0 & 0 & 1 \end{array} \right ) 
\eeq
with
\beq
\label{aldef}
\alpha \equiv \frac{m_2^2-m_1^2}{m_3^2-m_1^2} ~.
\eeq

In this limit, the distance $L$ from the source to the detector corresponding to $S(t',t)$ in Eq.~(\ref{smtx}) is
\beq
\label{Ldef}
L &=& t'-t ~.
\eeq
The time-evolution operator, Eq.~(\ref{smtx}), expressed in dimensionless variables is,
\beq
\label{smtx1}
S(L) &=& e^{-i  H_\nu (t' - t)} \nonumber \\
&=& e^{2i {\hat {\bar E}}^0_1 \Delta_L}  e^{-2i  {\hat {\bar H}}_\nu \Delta_L} ~,
\eeq
where ${\hat {\bar H}}_\nu $ is given in Eq.~(\ref{fulh1}), and where 
\beq
\label{Deldeff}
\Delta_L &\equiv& \frac{L(m^2_3-m^2_1)}{4E} ~.
\eeq
[The similar quantity $\Delta_L$ as defined in Ref.~\cite{hjk1} is exactly one-half of that appearing in Eq.~(\ref{Deldeff}).]


\subsection{Neutrino Mass Eigenvalues}
\label{diagH}

The neutrino mass eigenstates in a medium are solutions to the EV equation for ${\hat {\bar H}}_\nu$, Eq.~(\ref{eveq2}). 
In many familiar formulations~\cite{f,ahlo,JO} the full solutions, including both the eigenstates $|\nu_i>$ and EV's ${\hat {\bar E_i}}$, are required to find the neutrino oscillation probabilities.  

\subsubsection{Diagonalization of Neutrino Hamiltonian}

The energies ${\hat {\bar E_i}}$ are 
solutions of the cubic equation~\cite{KS}
\beq
\label{eqbarhatE1}
&&{\hat {\bar E_i}}^3 +a {\hat {\bar E_i}}^2 + b {\hat {\bar E_i}} +c   = 0 ~,
\eeq
where
\beq
\label{abcvals}
a &=& -(1+ \alpha + \hat A ) \nonumber \\
b &=& \alpha +  \hat A \cos^2{ \theta_{13}} +\hat A \alpha C^{(+)}_2 \nonumber \\ 
c &=& -\hat A \alpha \cos^2{ \theta_{12}} \cos^2{ \theta_{13}} ~.
\eeq
We have expressed $b$ in terms of a frequently recurring combination of mixing angles,
\beq
\label{Cbval}
C_2^{(\pm)} &\equiv& \cos^2{ \theta_{12}} \pm \sin^2{ \theta_{12}}  \sin^2{ \theta_{13}} ~.
\eeq
Note that the mass eigenstate energies  are independent of $\delta_{cp}$ and $\theta_{23}$ for both neutrinos and antineutrinos.

The solutions of Eq.~(\ref{eqbarhatE1}) are expressed conveniently in terms of the quantity $d$, 
\beq
\label{dalt}
d &=& \psi+ \sqrt{\psi^2-4\gamma^3} \nonumber \\\
\gamma &\equiv& a^2-3 b \nonumber \\
\psi &\equiv& a^3 - 27 c - 3 a \gamma
 ~.
\eeq
These solutions are real when
\beq
\label{erealr}
|d^{1/3}|^2 &=& 2^{2/3} \gamma ~ >  0 ~,
\eeq
which requires 
\beq
\label{psigam}
\psi^2 < 4 \gamma^3 ~,
\eeq
and, thus, that $d$ be complex. Because having real energies is required by Hermiticity of the neutrino Hamiltonian, Eqs.~(\ref{erealr},\ref{psigam}) amount to conditions on all parameter sets in terms of which $H_\nu$ is defined.

We find
\beq
\label{solneveq}
{\hat {\bar E_1}}  &=& -\frac{a}{3} - \frac{1 }{ 3 \cdot 2^{1/3} } ( \sqrt{3}  Im[ d^{1/3}] + Re[ d^{1/3}])  
\nonumber \\
{\hat {\bar E_2}} &=& -\frac{a}{3} + \frac{1 }{ 3 \cdot 2^{1/3} } (\sqrt{3} Im[ d^{1/3}] - Re[ d^{1/3}]) \nonumber \\
{\hat {\bar E_3}} &=& -\frac{a}{3} + \frac{2^{2/3} }{ 3} Re[ d^{1/3}]  ~.
\eeq
The masses are ordered so that $ m_3 > m_2 > m_1$. Because EV do not cross, $ {\hat {\bar E}}_3 > {\hat {\bar E}}_2 >{\hat {\bar E}}_1$ for all $|{\hat A}|$. 
A simple constraint among $ {\hat {\bar E}}_i$ is found from the trace of Eq.~(\ref{hamhat1}),
\beq
\label{trhamhat1}
Tr{\hat {\bar H}}_\nu &=& {\hat {\bar E_1}} + {\hat {\bar E_2}} + {\hat {\bar E_3}} \nonumber \\
&=& Tr {\hat {\bar H}}_{0v} + Tr {\hat   H}_1 \nonumber \\
&=&  1 + \alpha + {\hat A}  \equiv -a ~. 
\eeq

\subsubsection{Using neutrino mass eigenvalues in our Hamiltonian Formulation}

In our formulation, the entire dependence of the time evolution operator on the neutrino eigenvalues occurs through three eigenvalue combinations,
\beq
\label{SPD}
\Delta {\hat {\bar E}}_{\ell\ell'} &\equiv&   {\hat {\bar E_\ell}}  - {\hat {\bar E_{\ell'}}}  \nonumber \\
\Sigma_{\ell\ell'} &\equiv&  {\hat {\bar E_\ell}}  + {\hat {\bar E_{\ell'}}} \nonumber \\
\Pi_{\ell\ell'} &\equiv&   {\hat {\bar E_\ell}}  {\hat {\bar E_{\ell'}}}  ~,
\eeq
with $\ell > \ell'$ (and powers thereof). We denote such quantities using a bracket notation,  For example, 
\beq
\label{DelEdef}
\Delta {\hat {\bar E }}[1] &=& {\hat {\bar E}}_3 - {\hat {\bar E}}_2 \nonumber \\
\Delta {\hat {\bar E }}[2] &=& {\hat {\bar E}}_3 - {\hat {\bar E}}_1 \nonumber \\
\Delta {\hat {\bar  E }}[3] &=& {\hat {\bar E}}_2 - {\hat {\bar E}}_1 ~, 
\eeq
in the case of $\Delta {\hat {\bar E}} $.  We will generally use this bracket notation also for other quantities in our formulation that depend on two indices $(\ell,\ell')$, such as $\Sigma_{\ell\ell'} $ and $\Pi_{\ell\ell'} $.

An expression for $ \Sigma [\ell] $, 
\beq
\label{Sig1}
\Sigma[\ell] &=& -a  - {\hat {\bar E_\ell}} ~,
\eeq
follows from Eq.~(\ref{trhamhat1}).  An equivalent expression for $\Pi[\ell] $ in terms of ${\hat {\bar E_\ell}}$ is found by  subtracting Eq.~(\ref{eqbarhatE1}) for ${\hat {\bar E_\ell}} $  and that for ${\hat {\bar E_{\ell'}}} $ and dividing through by $ \Delta {\hat {\bar E}}_{\ell\ell'} $. We find
\beq
0 &=& ( {\hat  {\bar E_\ell}}^2 + {\hat {\bar E_\ell}} {\hat {\bar E_{\ell'}}} + {\hat {\bar E_{\ell'}}}^2 ) \nonumber \\
&+& a ( {\hat {\bar E_\ell}}  + {\hat {\bar E_{\ell'}}} ) + b \nonumber \\
&=& ( {\hat {\bar E_\ell}} + {\hat {\bar E_{\ell'}}} )^2 -  {\hat {\bar E_\ell}}  {\hat {\bar E_{\ell'}}} \nonumber \\
&+& a ( {\hat {\bar E_\ell}} + {\hat {\bar E_{\ell'}}} )  + b ~,
\eeq
giving
\beq
\Sigma[\ell]^2 -  \Pi[\ell]  +  a \Sigma[\ell]  + b &=& 0 ~.
\eeq
Then, using Eq.~(\ref{Sig1}),
\beq
\label{Pi1}
\Pi[\ell] &=& \Sigma_\ell  ( \Sigma_\ell +  a)  + b  \nonumber \\
&=& b + a {\hat {\bar E_\ell}} + {\hat {\bar E_\ell}}^2 ~.
\eeq

Finally, having observed that powers of the quantities given in Eq.~(\ref{SPD}) will appear in various expressions, we note that $\Pi[\ell]^p$ and $\Sigma[\ell]^q$ with $p \geq 2$ and $q \geq 3$ involve linear combinations of eigenvalues ${\hat {\bar E_\ell}}^n$ with powers $n \geq 3$.  Such terms  are equivalently represented by a linear combination of three terms, one  proportional to ${\hat {\bar E_\ell}}^2$, one proportional to ${\hat {\bar E_\ell}}$, and one independent of ${\hat {\bar E_\ell}}$, obtained  by using the equation of motion repeatedly.  We later make use of this fact to simplify various expressions.

\section{ The S-Matrix in Our Hamiltonian  Formulation}
\label{LFn}

The probability $\mathcal{P}(\nu_a  \rightarrow \nu_b)$ for neutrinos to oscillate from the initial state of flavor $a$ to a final state of flavor $b$ is found from the time-evolution operator $ S(t',t) $ as
\beq
\label{oscprob0}
\mathcal{P}(\nu_a  \rightarrow \nu_b) &=&   | S^{ab}(t',t)|^2 \nonumber \\
&\equiv& P^{ab}(t'-t) ~,
\eeq
where
\beq
\label{oscprob1}
|S^{ab}(t',t)|^2  &=&  | <\nu^0_{fb}| S(t',t) |\nu^0_{fa}> |^2  ~.
\eeq
We accordingly determine here $ P^{ab}(t'-t) $ from $ S(t',t) $ defined in Eq.~(\ref{smtx}). 

In this section we review the formulation of neutrino oscillations based on the Lagrange interpolation formula as used in Ref.~\cite{barg}. This formulation leads to  exact, closed-form expressions for the time-evolution operator and the partial oscillation probabilities that are linear combinations of spherical Bessel functions in the eigenvalue differences whose coefficients are polynomials in the neutrino CKM matrix elements, the neutrino mass differences squared, the strength of the neutrino interaction with matter, and the neutrino mass eigenvalues in matter.   We are led quite naturally to such expressions for all the partial oscillation probabilities in terms of these basic quantities.  The numerical results given later in this paper are based on this formulation.

\subsection{ Time-Evolution Operator }
\label{LF}

The overall phase in Eq.~(\ref{smtx1}) does not contribute to $| S^{ab}(t',t)|^2$, so for the purpose of calculating the oscillation probability, we may take 
\beq
\label{oscprob1}
S(L) &\to&    e^{-i  {\hat {\bar H}}_\nu \Delta_L} ~.
\eeq
Then, with neutrinos created and detected in flavor states, which are coherent linear combinations of the neutrino vacuum mass eigenstates given in Eq.~(\ref{MBS}),
\beq
|\nu^{0}_{fa}> &=& \sum_{j} U^*_{a j} |M(j)>  ~,
\eeq
we see that the mass eigenstate components of the flavor states contribute coherently to the time-evolution operator.  Thus,
\beq
\label{ampMb}
&&<\nu^0_{fb}| e^{-i  {\hat {\bar H}}_\nu \Delta_L} |\nu^0_{fa}> \nonumber \\
&=& <M(b)| U e^{-i  {\hat {\bar H}}_\nu \Delta_L} U^\dag |M(a)> ~.
\eeq
This coherence leads to the oscillation phenomenon.

The elegant formulae for $ S(L) \equiv e^{-i  {\hat {\bar H}}_\nu \Delta_L} $ are obtained from the Lagrange interpolation formula, Eqs.~(9,11) of Ref.~\cite{barg},
\beq
\label{lagrf}
U  e^{-i  {\hat {\bar H}}_\nu \Delta_L} U^{-1} &=& \sum_{\ell} F_\ell \exp^{-i {\hat {\bar E}}_\ell \Delta_L } ~,
\eeq
where $T = L = t' - t$ and 
\beq 
\label{Wdef}
F_\ell &\equiv& \Pi_{j\neq \ell} \frac{ U {\hat {\bar H}}_\nu U^{-1}  - {\bf 1} {\hat {\bar E}}_j }{ {\hat {\bar E}}_{\ell} - {\hat {\bar E}}_j} ~.
\eeq
For three neutrinos, the sum in Eq.~(\ref{lagrf}) runs over three values of $\ell$ and the product in Eq.~(\ref{Wdef}) over two values of $j$.

Using the convention that $O^{ab}$, written without parentheses enclosing $ab$, denotes the matrix elements of the operator $O$, 
\beq
O^{ab} &\equiv& <M(b)| O |M(a)> ~,
\eeq
the matrix elements $ F^{ab}_{\ell} $ of $ F_{\ell} $ given in Eq.~(\ref{Wdef})
may be compactly written 
\beq 
\label{Fabal1}
F^{ab}_{\ell} &=& \frac{ <M(b)| {\hat {\bar W}}[\ell] |M(a)> } { {\hat {\bar D}}[\ell] }  ~,
\eeq
where~\cite{barg},
\beq 
\label{Wdef3}
{\hat {\bar W}}[1] &\equiv& (U {\hat {\bar H}}_\nu U^{-1} - {\bf 1}{\hat {\bar E}}_3 ) (U {\hat {\bar H}}_\nu U^{-1} - {\bf 1}{\hat {\bar E}}_2 ) \nonumber \\
{\hat {\bar W}}[2] &\equiv&   (U {\hat {\bar H}}_\nu U^{-1} - {\bf 1}{\hat {\bar E}}_3 ) (U {\hat {\bar H}}_\nu U^\dag - {\bf 1}{\hat {\bar E}}_1 )\nonumber \\
{\hat {\bar W}}[3] &\equiv& (U {\hat {\bar H}}_\nu U^{-1} - {\bf 1}{\hat {\bar E}}_2 ) \nonumber \\
&\times& (U {\hat {\bar H}}_\nu U^\dag - {\bf 1}{\hat {\bar E}}_1 ) 
\eeq
and
\beq
\label{Dal1}
{\hat {\bar D}}[1] &=& ({\hat {\bar E}_{3}} - {\hat {\bar E}_1} )({\hat {\bar E}_{2}} - {\hat {\bar E}_1})  \nonumber \\
{\hat {\bar D}}[2] &=&  ({\hat {\bar E}_{1}} - {\hat {\bar E}_2})  ({\hat {\bar E}_{3}} - {\hat {\bar E}_2}) \nonumber \\
{\hat {\bar D}}[3] &=&  ({\hat {\bar E}_{1}} - {\hat {\bar E}_3}) ({\hat {\bar E}_{2}} - {\hat {\bar E}_3}) ~. 
\eeq
Equations~(\ref{Wdef3},\ref{Dal1}) use the same bracket notation introduced in Eq.~(\ref{Deldeff}).
The result in Eqs.~(\ref{ampMb},\ref{lagrf},\ref{Wdef}) is immediately verified to be correct by inserting a complete set of intermediate eigenstates of $H_\nu$ in Eq.~(\ref{Wdef3}).

It follows from the unitarity of $U$ that $ {\hat {\bar W}}[\ell] $ is Hermitian, 
\beq
\label{hermitp}
{\hat {\bar W}}[\ell]^\dagger &=&  {\hat {\bar W}}[\ell] 
\eeq
and that the two factors in Eqs.~(\ref{Wdef3}) commute with each other. We find from Eq.~(\ref{Wdef3}) that 
\beq
\label{TrW1}
Tr  {\hat {\bar W}}[\ell]  &=&  {\hat {\bar D}} [\ell]
~.
\eeq
Equation~(\ref{hermitp}) establishes the reflection symmetry 
\beq
\label{sFabal1}
F^{ab*}_{\ell} &=& F^{ba}_{\ell} ~.
\eeq

Explicit expressions for ${\hat {\bar W}}[\ell]$ are easily found in terms of $H_\nu$.  The entire dependence of ${\hat {\bar W}}[\ell]$ on $\delta_{cp}$ occurs through three operators independent of $\delta_{cp}$, $W_0[\ell]$,  $W_{cos}[\ell]$ and $W_{sin}[\ell]$,
\beq
\label{PstruA}
{\hat {\bar W}}[\ell] &=& {\hat {\bar W}}_0[\ell] + \cos{\delta_{cp}} {\hat {\bar W}}_{cos}[\ell] \nonumber \\
&+& i \sin{\delta_{cp}} {\hat {\bar W}}_{sin}[\ell] ~,
\eeq
with ${\hat {\bar W}}_0[\ell]$,  ${\hat {\bar W}}_{cos}[\ell]$ and ${\hat {\bar W}}_{sin}[\ell]$ real and independent of $\delta_{cp}$. Details are given in Appendix~\ref{a:LG}.

\subsection{Total Oscillation Probability  }

Expressions for $\mathcal{P}(\nu_a  \rightarrow \nu_b)$ may be obtained directly from $S(L)$,
\beq
\label{oscprob0} 
\mathcal{P}(\nu_a  \rightarrow \nu_b) &=&   | S^{ab}(t',t)|^2 \nonumber \\
&=&  Re[S^{ab}(L)]^2 + Im[S^{ab}(L)]^2~.
\eeq
Convenient expressions for $Re[S^{ab}(L)]$ and $Im[S^{ab}(L)]$ defined by Eq.~(\ref{lagrf}) are presented in Appendix~\ref{a:LG}. 
In our Hamiltonian formulation, the dependence of $ S(T)$ on the CP violating phase $\delta_{cp}$ is very simple and follows from Eqs.~(\ref{lagrf},\ref{PstruA}) noting that $ F^{ab}_{\ell} = {\hat {\bar W}}^{ab}[\ell] / {\hat {\bar D}}[\ell] $, Eq.~(\ref{Fabal1}). We thus find,  
\beq
\label{lagrf6}
Re[ S^{ab}(t',t) ] &=&  \delta_{ab} -  2 \sum_{\ell}  
\frac{ {\hat {\bar W}}^{ab}_0[\ell]} { {\hat {\bar D}}[\ell] }\sin^2{ {\hat{\bar E}}_{\ell} \Delta_L  } \nonumber \\
&-&  2 \cos{\delta_{cp}} \sum_{\ell}  
\frac{ {\hat {\bar W}}^{ab}_{cos}[\ell] } { {\hat {\bar D}}[\ell] }\sin^2{ {\hat{\bar E}}_{\ell} \Delta_L  } \nonumber \\
&+& \sin\delta_{cp} \sum_{\ell}  \frac{ {\hat {\bar W}}^{ab}_{sin}[\ell]  } { {\hat {\bar D}}[\ell] }\sin{ 2{\hat{\bar E}}_{\ell} \Delta_L  }    ~,
\eeq
and
\beq
Im[ S^{ab}(t',t) ] &=&  -  2 \sin\delta_{cp} \sum_{\ell}  
\frac{  {\hat {\bar W}}^{ab}_{sin}[\ell] } { {\hat {\bar D}}[\ell] }  \sin^2{ {\hat{\bar E}}_{\ell} \Delta_L  } \nonumber \\
&-& \sum_{\ell} \frac{  {\hat {\bar W}}^{ab}_0[\ell] } { {\hat {\bar D}}[\ell] }  \sin{ 2{\hat{\bar E}}_{\ell} \Delta_L  } \nonumber \\
&-& \cos{\delta_{cp}} \sum_{\ell} \frac{  {\hat {\bar W}}^{ab}_{cos}[\ell] } { {\hat {\bar D}}[\ell] }  \sin{ 2{\hat{\bar E}}_{\ell} \Delta_L  } ~,
\eeq
where $\Delta_L$ is defined in Eq.~(\ref{Deldeff}).
Approximate expressions for $ \mathcal{P}(\nu_a  \rightarrow \nu_b) $ in terms of the parameters of $ H_\nu$ were obtained from $S(L)$ in Refs.~\cite{ahlo,JO} by an expansion in $\sin\theta_{13}$.

\subsection{Partial Oscillation Probabilities   }
\label{POPL}

Using somewhat different techniques, the oscillation probability  may be expressed through a set of functions that express  how 
$\mathcal{P}(\nu_a  \rightarrow \nu_b) \equiv P^{ab}$  depends on the CP violating phase $\delta_{cp}$~\cite{f}. In our Hamiltonian formulation there are four such terms,
\beq
\label{Pmmemu0}
P^{ab} &=& \delta(a,b) +  P_0^{ab} + P_{\sin\delta}^{ab} + P_{\cos\delta}^{ab} \nonumber \\
&+& P_{\cos^2\delta}^{ab} ~,
\eeq
with $P_{\sin\delta}^{ab}$ linear in $\sin{\delta_{cp}}$,
$P_{\cos\delta}^{ab}$ linear in $\cos{\delta_{cp}}$,
$P_{\cos^2\delta}^{ab}$ quadratic in $\cos{\delta_{cp}}$,
and $P_0^{ab}$ independent of $\delta_{cp}$. Although only the overall oscillation probability is a true probability, guaranteed to be strictly positive everywhere, we find it convenient to refer to these four terms as ``partial oscillation probabilities." Approximate expressions for the partial oscillation probabilities expanded in the small parameter $\alpha$ of the SNM in Ref.~\cite{f}.

Obtaining expressions for the partial oscillation probabilities from the time-evolution operator requires additional analysis, given in Appendix~\ref{PartOscProb}.  In terms of spherical Bessel functions, we find there,
\beq
\label{POPLdef1a} 
P_{sin\delta}^{ab} (\Delta_L,\hat A)   &=&  \sin { \delta_{cp} }  \frac{4 \Delta_L }  {{\hat D}} \sum_\ell (-1)^\ell w_{sin}^{ab} [\ell] \nonumber \\
&\times& j_0 ( 2 {\hat \Delta}[\ell] ) ~,
\eeq
where $ w_{sin}^{ab} [\ell] $, and therefore $ P_{sin\delta}^{ab} $, is  anti-symmetric under $a\leftrightarrow b$.  
The other three partial oscillation probabilities are individually symmetric under $a\leftrightarrow b$.  We find
\beq
\label{POPLdef1b}
P_{cos\delta}^{ab}(\Delta_L,\hat A)   &=& - \cos {\delta_{cp} } \frac{ 4 \Delta_L^2} {  {\hat {\bar D}} }  \sum_\ell (-1)^\ell  w_{cos}^{ab} [\ell]  \nonumber \\
&\times& \Delta \hat {\bar E}}[\ell] j_0^2{ ( {\hat \Delta}[\ell] ) \nonumber \\
P_{cos^2\delta}^{ab}(\Delta_L,\hat A)   &=& - \cos^2 { \delta_{cp} } \frac{ 4 \Delta_L^2} {  {\hat {\bar D}} }   \sum_\ell (-1)^\ell  w_{cos^2}^{ab} [\ell]  \nonumber \\
&\times& \Delta \hat {\bar E}}[\ell] j_0^2{ ( {\hat \Delta}[\ell] ) \nonumber \\
P_{0}^{ab}(\Delta_L,\hat A)   &=& - \frac{ 4 \Delta_L^2} {  {\hat {\bar D}} }   \sum_\ell (-1)^\ell  w_{0}^{ab} [\ell]  \nonumber \\
&\times& \Delta \hat {\bar E}}[\ell] j_0^2{ ( {\hat \Delta}[\ell] )  ~,
\eeq
where all sums run over $\ell=1~,2~,3$, ${\hat \Delta}[\ell]  $ is defined as 
\beq
\label{hatdeldef}
{\hat \Delta}[\ell]  &\equiv& \Delta {\hat{\bar E}}[\ell] \Delta_L ~,
\eeq 
with $\Delta {\hat {\bar E}[\ell]} $ defined in Eq.~(\ref{DelEdef}), ${\hat {\bar D} }$ is defined as
\beq
\label{Dhatbardef}
{\hat {\bar D} } &\equiv& \Delta {\hat {\bar E}[1]} \Delta {\hat {\bar E}[2]} \Delta {\hat {\bar E}[3]} ~,
\eeq
and the matrix elements $ w_{i}^{ab} [\ell]  $ are given in terms of the mixing angles and ${\hat A}$ in Appendix~\ref{PartOscProb}.  
Since we order the energies so that $ {\hat {\bar E}}_3 > {\hat {\bar E}}_2 >{\hat {\bar E}}_1 $, $\Delta {\hat {\bar E}[\ell]}$ as well as  ${\hat {\bar D} }$ are all positive.  

We begin our derivation with
the expression for the oscillation probability  written in terms of $S(T)$, Eq.~(\ref{lagrf}), 
\beq
\label{oscprob}
&&\mathcal{P}(\nu_a  \rightarrow \nu_b) \equiv P^{ab} \nonumber \\
&=&  |<M(b)| U e^{-iH_{\nu}(t' -t)} U^\dag |M(a)> |^2 \nonumber \\
&=& \sum_{\ell \ell' } F^{ab}_{\ell  \ell' }  \exp^{-i ( {\bar E}_{\ell} - {\bar E}_{\ell'} )L}  ~.
\eeq
Here $ F^{ab}_{\ell  \ell'} $ is defined as
\beq
\label{Pmab1}
F^{ab}_{\ell  \ell'} &\equiv& F^{ab}_{\ell} F^{ab*}_{\ell'} = \frac{ w^{ab}_{\ell\ell'}  } { {\hat {\bar D}}[\ell]   {\hat {\bar D}}[\ell']  }
\eeq
with ${\hat {\bar D}}[\ell] $ given in Eq.~(\ref{Dal1}), and 
\beq
\label{Pmab1wt}
w^{ab}_{\ell \ell'} &\equiv&   <M(b)| {\hat {\bar W}}[\ell] |M(a)> \nonumber \\
&\times& <M(b)|{\hat {\bar W}}[\ell'] |M(a)>^*  ~.
\eeq
All results needed for determining exact, closed-form  expressions for the partial oscillation probabilities are found in Appendix~\ref{PartOscProb}.

As we explain in Appendix~\ref{PartOscProb}, $ w^{ab}_{\ell \ell'}$ may be expressed through four operators,  
\beq
\label{a:Pmabs3}
w^{ab}_{\ell\ell'} &=& w_{0\ell\ell'}^{ab} + \cos{\delta_{cp}} w_{cos~\ell\ell'}^{ab} + \cos^2{\delta_{cp}} \nonumber \\
&\times& w_{cos^2\ell\ell'}^{ab} 
+ i \sin{\delta_{cp}} w_{sin\ell\ell'}^{ab} ~. 
\eeq
found from the decomposition of ${\hat {\bar W}}[\ell] $ given in Eq.~(\ref{PstruA}).  The quantities $ w_{i}^{ab} [\ell]  $ appearing in Eqs.~(\ref{POPLdef1a},\ref{POPLdef1b}) are the same as $ w_{i\ell\ell'}^{ab} $ written in the streamlined notation, 
\beq
w_{i}^{ab} [1]  &=& w_{i\ell\ell'}^{ab} ~~{\rm for} ~~(\ell,\ell')=(3,2) \nonumber \\
w_{i}^{ab} [2]  &=& w_{i\ell\ell'}^{ab}~~{\rm for} ~~(\ell,\ell')=(3,1)  \nonumber \\
w_{i}^{ab} [3]  &=& w_{i\ell\ell'}^{ab}~~{\rm for} ~~(\ell,\ell')=(2,1)  ~,
\eeq
which takes advantage of $\ell > \ell'$. 

An analytic expression for $ w^{ab}_{sin}[\ell]$,
\beq
\label{wsineva2c0}
w^{ab} _{sin}[\ell] &=&  K  \sin{2\theta_{23}} \Delta {\hat{\bar E}}[\ell] \epsilon^{ab}_{\sin} ~,
\eeq
follows directly from Eq.~(\ref{PstruA}).  Here,  $\epsilon_{\sin}$ is the anti-symmetric matrix 
\beq
\label{epsidef}
\epsilon_{sin} &\equiv& \left( \begin{array}{ccc} 0 &  1 &  -1  \\  -1   & 0 & 1 \\     1 &  -1    & 0 \end{array} \right) 
\eeq
and
\beq
\label{Kdef}
K &=& -\frac{\alpha(1-\alpha) }{8} \nonumber \\
&\times& \cos{\theta_{13}} \sin{2\theta_{12}} \sin{2\theta_{13}} ~.
\eeq
Equation~(\ref{wsineva2c0}) is one of the more striking results. Analytic formulae for the other  $ w_{i}^{ab}[\ell] $ follow from Eq.~(\ref{Fabal1}).  These are all given in Appendix~\ref{PartOscProb}, where they are expressed in terms of the parameters specifying ${\hat {\bar H}}_{0v}$ and ${\hat H}_1 = U^{-1} {\hat V} U$. These are the same parameters defining SNM. 

It follows from Eq.~(\ref{wsineva2c0}) that the  coefficients of $ \Delta {\hat{\bar E}}[\ell] j_0 ( 2 {\hat \Delta}[\ell] ) $ in Eq.~(\ref{POPLdef1a})
are all proportional, leading to a simple expression for $P_{sin\delta}^{ab}$,
\beq
\label{a:ppsin1c2}
P_{sin\delta}^{ab}(\Delta_L,\hat A)  
&=& \Delta_L^3   \alpha (1-\alpha) \sin {\delta_{cp} }  \epsilon_{sin}^{ab}  \nonumber \\
&\times& \cos{\theta_{13}} \sin{ 2 \theta_{12}} \sin{2\theta_{13}} \sin{2\theta_{23}} \nonumber \\
&\times& j_0  ({\hat \Delta}[3] )       j_0 ( {\hat \Delta}[2] )    j_0 (  {\hat \Delta}[1] ) ~.
\eeq
Note that $P^{ab}_{\sin\delta}$ is antisymmetric under $a\leftrightarrow b$.

Analytic expressions for all other partial oscillation probabilities follow from Eq.~(\ref{oscprob}) using Eq.~(\ref{Pmab1}), which expresses $ F^{ab}_{\ell  \ell'} $ in terms of $w^{ab}_{\ell\ell'} $.
In this fashion, these partial oscillation probabilities are also expressed in terms of the parameters of the SNM and the neutrino eigenvalues, ${\hat{\bar E}}_\ell$. 

The usefulness of the partial oscillation probabilities can be seen as follows.  
It is a general result that the exchange of initial and final states in the oscillation probability or neutrinos (antineutrinos) is equivalent to letting $\delta_{cp}\rightarrow -\delta_{cp}$. Thus, the result for the inverse reaction $\mathcal{P}(\nu_b  \rightarrow \nu_a)$ is found by exchanging $(a,b)$ in Eq.~(\ref{Pmmemu0}).  Since  $ P_{\sin\delta}^{ab} $ is antisymmetric under the exchange of $(a,b)$, and $ P_0^{ab}$, $P_{\cos\delta}^{ab}$ and 
$P_{\cos^2\delta}^{ab}$ symmetric, 
it follows that $P^{ba}$ is given by
\beq
\label{Pmmume0}
\mathcal{P}(\nu_b  \rightarrow \nu_a) &=& \delta(a,b) +  P_0^{ab} - P_{\sin\delta}^{ab} + P_{\cos\delta}^{ab} \nonumber \\
&+& P_{\cos^2\delta}^{ab} ~.
\eeq

In analogy to Eq.~(\ref{Pmmemu0}), we may express the oscillation probability for antineutrinos as
\beq
\label{Pbarmemu0}
&&\mathcal{P}({\bar \nu}_a  \rightarrow {\bar \nu}_b) \equiv  {\bar P}^{ab} \nonumber \\
&=& \delta(a,b) + { \bar P}_0^{ab} + {\bar P}_{\sin\delta}^{ab} + {\bar P}_{\cos\delta}^{ab} + {\bar P}_{\cos^2\delta}^{ab}
  ~,
\eeq
where the bared probabilities for anti neutrinos are obtained from the unbarred for neutrinos by replacing $\delta_{cp}\rightarrow -\delta_{cp}$ and ${\hat A} \rightarrow - {\hat A}$. Because the energies of antineutrinos are different from those of the neutrinos in matter, we can expect $P^{ab} \neq {\bar P}^{ab}$ in this situation.

Again applying the rule that exchange of initial and final states is accomplished by letting $\delta_{cp}\rightarrow -\delta_{cp}$, the oscillation probability $\mathcal{P}({\bar \nu}_b  \rightarrow {\bar \nu}_a)$ is expressed in terms of the same four quantities,
\beq
\label{Pbarmmue0}
&&\mathcal{P}({\bar \nu}_b  \rightarrow {\bar \nu}_a) \nonumber \\
&=& \delta(a,b) +  {\bar P}_0^ {ab} -  {\bar P}_{\sin\delta}^{ab}   + {\bar P}_{\cos\delta}^{ab}  + {\bar P}_{\cos^2\delta}^{ab}
 ~.
\eeq

It is worth noting that the entire dependence of the oscillation probabilities given in Eqs.~(\ref{POPLdef1a},\ref{POPLdef1b})  on the neutrino beam energy $E$, the baseline $L$, and the medium properties occurs through the variables $\Delta_L$ and ${\hat A}$ defined in Eqs.~(\ref{Deldeff},\ref{hatAdef}), respectively. Since we will be most interested in how the neutrino oscillation probability depends on the beam energy, baseline, and medium properties, the partial oscillation probabilities have been expressed as functions of $\Delta_L$ and ${\hat A}$. 

Because the vacuum  result is also easily obtained by less sophisticated arguments, the vacuum limit provides an opportunity to verify our Hamiltonian formulation in a well-known special case. 

\subsection{ Special Cases } 
\label{LPVac}

We next examine  special cases chosen to highlight specific aspects of  our formulation.  Because these  cases  are  well known in other contexts, they provide useful cross checks. We first consider  two-flavor mixing and then the vacuum limit.

\subsubsection{ Two Flavor Mixing}

Expressions for two-neutrino oscillations are easily obtained from the Hamiltonian,
\beq
\label{2fH}
{\hat {\bar H}}_\nu = {\hat {\bar H}}_{0\nu}+ U^{-1} {\hat {\bar H}}_{1\nu} U ~,
\eeq
where 
\beq
{\hat {\bar H}}_{0 \nu} &\equiv& \left( \begin{array}{ccc} 0 &  0 &    \\  0  & 1  \end{array} \right) ~,
\eeq
\beq
{\hat {\bar H}}_{1\nu} &\equiv&   \left( \begin{array}{ccc} {\hat A} &  0   \\  0  & 0  \end{array} \right) ~.
\eeq
The standard mixing matrix $U$ is
\beq
\label{UtwoF}
U &\equiv& \left( \begin{array}{ccc} \cos\theta & \sin\theta \\ -\sin\theta & \cos\theta   \end{array} \right ) ~.
\eeq 
The neutrino Hamiltonian is easily diagonalized to find the medium-modified  two-flavor neutrino eigenvalues,
\beq
\label{EVtwo}
{\hat {\bar E_2}} &=& \frac{1}{2}\left(1+\hat A + \phi \right) \nonumber \\
{\hat {\bar E_1}}  &=& \frac{1}{2}\left(1+\hat A - \phi \right)  ~,
\eeq
where $\phi =  \sqrt{1 + {\hat A}^2-2 {\hat A} \cos{2 \theta}}$.

Expressed in dimensionless variables, the time-evolution operator $S^{ab}$ for the transition $a \to b$ becomes
\beq
S^{a b }&=&  <M(b)| U \exp^{-2 i {\hat {\bar H}}_\nu \Delta_L}U^{-1}    |M(a)> ~,
\eeq
with  $\Delta_L$ given in  Eq.~(\ref{Deldeff}).  The two-flavor oscillation probability $P^{12}_{2f}(\Delta_L,\hat A) $  is then
\beq
\label{P2f}
P^{12}_{2 f}(\Delta_L,\hat A)  &=& |S_{2f}^{12}|^2 \nonumber \\
&=& \frac{ \sin^2 2 \theta }{ \phi^2 }   \sin^2\Delta_L \phi ~,
\eeq
which we will next compare to our Hamiltonian formulation.  To find $S_{2f}^{1 2}=<M(2)|S[L]|M(1)>$, it is easy to overlook that the  $2\times 2$ matrix $H_\nu$ in Eq.~(\ref{2fH}) is not diagonal in the flavor basis .  

The oscillation probability in the three neutrino mixing of our Hamiltonian formulation is matched to the two-flavor case just discussed by setting two of the mixing angles in Eq.~(\ref{Ubase}) to zero, say $\theta_{13}\to 0$ and $\theta_{23}\to 0$ and identifying $\theta$ with $\theta_{12}$.  Accordingly, we find,
\beq
\label{UtwoF}
U &\equiv& \left( \begin{array}{ccc} \cos\theta & \sin\theta & 0 \\ -\sin\theta & \cos\theta  & 0  \\  0 & 0 & 1  \end{array} \right ) ~.
\eeq 
This is easily recognized as a two-flavor mixing matrix by noting that one of the three neutrinos does not mix with the other two.
We see that it not only  depends on just one mixing angle $\theta$,  but also that the dependence on the CP violating phase $\delta$ has dropped out. 

Natural choices for a $3 \times 3$  two-flavor neutrino vacuum Hamiltonian and  interaction are,
\beq
{\hat {\bar H}}_{0 \nu} &\equiv& \left( \begin{array}{ccc} 0 &  0 &  0  \\  0  & 1 & 0 \\     0 &  0    & 0 \end{array} \right) ~,
\eeq
and
\beq
{\hat {\bar H}}_1 &\equiv&   \left( \begin{array}{ccc} {\hat A} &  0 &  0  \\  0  & 0 & 0 \\     0 &  0    & 0 \end{array} \right) ~.
\eeq

When the $3\times 3$ neutrino Hamiltonian  ${\hat {\bar H}}_\nu ={\hat {\bar  H}}_{0\nu}+ U^{-1} {\hat {\bar H}}_{1\nu} U$ is diagonalized, two of the three medium-modified neutrino eigenvalues,
\beq
\label{EVtwo}
{\hat {\bar E_3}} &=& \frac{1}{2}\left(1+\hat A + \phi \right) \nonumber \\
{\hat {\bar E_2}}  &=& \frac{1}{2}\left(1+\hat A -\phi \right) \nonumber \\
{\hat {\bar E_1}} &=& 0 ~.
\eeq
are identical to those found above in the two-flavor case.  Note that the eigenvalues have no branch points for ${\hat A}$ on the real axis. 
From these eigenvalues, we find the differences,
\beq
\Delta {\hat {\bar E}} [3] &=& {\hat {\bar E_2}} -{\hat {\bar E_1}} \nonumber \\
&=&  \frac{1}{2}\left(1+\hat A -\phi \right)  \nonumber \\
\Delta {\hat {\bar E}} [2] &=& {\hat {\bar E_3}} -{\hat {\bar E_1}} \nonumber \\
&=&  \frac{1}{2}\left(1+\hat A + \phi \right) \nonumber \\ 
\Delta {\hat {\bar E}} [1] &=& {\hat {\bar E_3}} -{\hat {\bar E_2}} = \phi  ~.
\eeq

Finally,  consider the oscillation probability in our Hamiltonian formulation.  Because $P^{ab}_{\sin\delta}$, $P^{ab}_{\cos\delta}$, and $P^{ab}_{\cos^2\delta}$ are proportional to $\sin 2 \theta_{23}$ (and $\theta_{23}$ has been set to $0$),  these terms do not contribute for two-flavor mixing. Thus, 
\beq
P^{ab}(\Delta_L,\hat A)&=&  \delta(a,b)+P^{ab}_0(\Delta_L,\hat A)
\eeq
with  $P^{ab}_0(\Delta_L,\hat A)$ taken from  Eq.~(\ref{POPLdef1b}).
We find after  a straightforward calculation,
\beq
w_0^{11} [\ell] &=& -w_0^{12}[\ell] \nonumber \\
w_0^{21} [\ell] &=& w_0^{12} [\ell]  \nonumber \\
w_0^{22} [\ell] &=& -w_0^{12}[\ell] ~,
\eeq
where,
\beq
w_0^{12} [\ell]  &=& w^{(12)}_{02}  ({\hat A} \cos^2\theta - (1+{\hat A} )  {\hat {\bar E}}_\ell  \nonumber \\
&+&  {\hat {\bar E}}_\ell ^2) ~.
\eeq

Taking  ${\hat {\bar E}}_\ell$ from Eq.~(\ref{EVtwo}), we evaluate 
\beq
{\hat A} \cos^2\theta - (1+{\hat A} )  {\hat {\bar E}}_\ell  +  {\hat {\bar E}}_\ell ^2  ~,
\eeq
to find
\beq
w_0^{12} [1] = w_0^{12} [2] = 0 ~,
\eeq
and 
\beq
w_0^{12} [3] &=& \hat A \cos^2\theta w^{(12)}_{0,2} \nonumber \\
&=& \frac{\hat A}{4}  \cos^2\theta  \sin^2 2\theta ~.
\eeq

Thus, for all two-flavor  transitions, 
\beq
\label{L2f}
P^{ab}(\Delta_L,\hat A) &=& \delta(a,b) + \frac{4 \Delta_L^2}{\hat D}    w_0^{ab} [3] \Delta {\hat {\bar E[3]}} \nonumber \\
&\times&  j_0^2(\Delta {\hat {\bar E[3]}} \Delta_L) ~,
\eeq
where   $\hat D$ is  from Eq.~(\ref{hatDval}),
\beq
\label{hatDval}
\hat D &\equiv& \Delta {\hat {\bar E}} [1] \Delta {\hat {\bar E}} [2] \Delta {\hat {\bar E}} [3]  \nonumber \\
 &=&    \frac{\Delta {\hat {\bar E}} [1] }{4} ((1+\hat A)^2 - 1 - {\hat A}^2 +2  {\hat A} \cos{2 \theta}  ) \nonumber \\
&=& \hat A \cos^2\theta \Delta {\hat {\bar E}} [1] ~.
\eeq

To compare our result to  to the well-known two-flavor oscillation probability given in Eq.~(\ref{P2f}), we evaluate Eq.~(\ref{L2f}) for $1 \to 2$  transitions obtaining,
\beq
P^{12}(\Delta_L,\hat A) &=&   \frac{\sin^2 2\theta  }{ \phi^2}    \sin^2  \Delta_L  \phi   ~.
\eeq
As expected, this is in complete agreement with Eq.~(\ref{P2f}).

\subsubsection{ Vacuum Oscillation Probability}

In our Hamiltonian formulation, an expression for the  time-evolution operator  in the vacuum limit $\hat A \to 0$ is found  from Eqs.~(\ref{ampMb}), (\ref{lagrf}), (\ref{Fabal1}), and (\ref{Dal1}),
\beq
\label{amp00L0} 
&&<M(b)|S^0(t',t)|M(a)> \equiv S^{0ab}(t'-t) \nonumber \\
&=& \sum_{\ell}  \frac{<M(b)| {\hat {\bar W^0}}[\ell] |M(a)> }{ {\hat {\bar D^0}}[\ell] } \exp^{-i \bar E^0_\ell \Delta_L }  ~.
\eeq
Here,  $ \bar E^0_\ell $ and  ${\hat {\bar D^0}}[\ell]$  are the vacuum values of $ \bar E_\ell $ and  ${\hat {\bar D}}[\ell]$, respectively.  In the vacuum, of course, there is no distinction between the energy differences for neutrinos and anti-neutrinos. 

Evaluating Eq.~(\ref{amp00L0}) by inserting a complete set of states intermediate states  $|n><n|$ inside ${\hat {\bar W}}^0[\ell]$, we arrive at
 
\begin{widetext}

\beq
<M(b)|S^0(t',t)|M(a)> &=&  \sum_{\ell } \sum_{n} U_{bn}  \frac{  ( { \hat {\bar E}}^0_n - {\hat {\bar E}}^0_{a })(  {\hat {\bar E}}^0_n -  {\hat {\bar E}}^0_{b} ) }{ {\hat {\bar D^0}}[\ell] } U_{na}^*  e^ { -i {\bar E}^0_\ell (t' - t) } ~.
\eeq

\end{widetext}
The indices  $(a, b, n)$ run over all permutations of the three integers $(1,2,3)$, from which it follows that $( { \hat {\bar E}}^0_n - {\hat {\bar E}}^0_{a[\ell] })(  {\hat {\bar E}}^0_n -  {\hat {\bar E}}^0_{b[\ell]} ) \equiv  {\hat {\bar D^0}}[\ell]$, and, consequently,
\beq
\label{amp00L}
&&<M(b)|S^0(t',t)|M(a)> \nonumber \\
&=& \sum_n U_{bn} e^{ -i E^0_n (t'-t)  } U_{na}^* ~.
\eeq
This  well-known vacuum limit also follows directly from more elementary considerations, using Eqs.~(\ref{tunif},\ref{Fdef}),
\beq
\label{amp00S}
&&  <\nu^0_{fb}|e^{-i H_{0v} (t' - t)}|\nu^0_{fa}> \nonumber \\
&= & \sum_n U_{bn} e^{ -i E^0_n (t'-t)  } U_{na}^* ~. 
\eeq
The equality of Eqs.~(\ref{amp00L},\ref{amp00S})  verifies our exact Hamiltonian formulation in the vacuum limit.  

Finally, we find that $F^{ab}_{\ell\ell'} $ appearing Eq.~(\ref{oscprob}) takes a familiar form in the vacuum.  Using Eq.~(\ref{amp00L}), or Eq.~(\ref{amp00S}), 

\begin{widetext}

\beq
\label{P0ab}
&& \mathcal{P}(\nu^0_a  \rightarrow \nu^0_b) = \left(\sum_\ell U_{b\ell}   e^{-i {\bar E^0_\ell} (t' -t)} U^*_{\ell a} \right) \left(\sum_{\ell'} U_{b\ell'}  e^{-i {\bar E^0_{\ell'} } (t' -t)} U^*_{\ell' a}\right)^* ~. 
\eeq
Expressing  this in terms of   Freund's $J^{ab}_{\ell\ell'}$~\cite{f},
\beq
J^{ab}_{\ell\ell'} &\equiv& U^*_{b\ell} U_{a\ell}  U^*_{a\ell'}  U_{b\ell'} ~, 
\eeq
and using Eq.~(\ref{Evac1}),
\beq
\label{Delhat0}
{\bar E}^0_{\ell} - {\bar E}^0_{\ell'} &=& \frac{m_\ell^2-m_{\ell'}^2}{2E} ~,
\eeq
we find that Eq.~(\ref{P0ab}) may be written, 
\beq
\label{probfreund}
&& \mathcal{P}(\nu^0_a  \rightarrow \nu^0_b) = \delta_{ab} -  4\sum_{i>j} Re J^{ab}_{ij} \sin^2{ ( ( {\bar E}^0_{\ell} - {\bar E}^0_{\ell'} )L )} - 2 \sum_{i>j} Im J^{ab}_{ij} \sin ( {2 ( {\bar E}^0_{\ell} - {\bar E}^0_{\ell'} )L ) } ~.
\eeq

\end{widetext}
Comparing Eqs.~(\ref{probfreund})  and (\ref{oscprob}), it is clear  that $F^{ab}_{\ell\ell'} \rightarrow J^{ab*}_{\ell\ell'}$ in the vacuum limit.

\section{The Standard Neutrino Model}

We adopt the Standard Neutrino Model~\cite{ISS} as our description of neutrino physics. The parameters defining the model include a (dimensionless) interaction strength ${\hat A}$ of neutrinos and anti-neutrinos with matter, the three neutrino mass differences, the three mixing angles, and the $CP$-violating phase. Most of the parameters of the SNM are consistent with global fits to neutrino oscillation data with relatively good precision. 

The neutrino mass differences of the SNM are taken to be~\cite{hjk1,khj2,hjk1a,khj1E,kiss1}  
\beq
m^2_2-m^2_1 &\equiv& \delta m_{21}^2 \nonumber \\
&=& 7.6\times 10^{-5} ~{\rm eV}^2 \nonumber \\
m^2_3-m^2_1 &\equiv& \delta m_{31}^2 \nonumber \\
&=& 2.4\times 10^{-3} ~{\rm eV}^2 ~, 
\eeq
corresponding to
\beq
\alpha &\equiv& \frac{\delta m_{21}^2}{\delta m_{31}^2} \nonumber \\
&=& 3.17\times 10^{-2} ~.
\eeq 
In Ref.~\cite{hjk1,khj2,hjk1a,khj1E,kiss1}, $\delta \equiv (m_2^2 - m_1^2)/(2E) $ and $\Delta = (m_3^2 - m_1^2)/(2E) $ whereas in Ref.~\cite{f}, ${\hat \Delta} = (m_3^2 - m_1^2)L/(4E)$ and $\Delta  = m_3^2 - m_1^2$.

The mixing angle $\theta_{23}$, 
\beq
\theta_{23}&=& \pi/4  ~,
\eeq
is the best-fit value from Ref.~\cite{Dav}, and $\theta_{12}$,
\beq
\theta_{12}&=& \pi/5.4 ~,
\eeq
is consistent with the recent analysis of Ref.~\cite{Gon}. 
The mixing angle $\theta_{13}$ is known to be small ($\theta_{13}<0.18$ at the 95\% confidence level) but its precise value is uncertain.  A very recent result from the Daya Bay project~\cite{DB}  is $\sin{\theta_{13}} \approx 0.15$, which we adopt to determine our value for $\theta_{13}$,
\beq
\theta_{13}&=& 0.151  ~.
\eeq
This fixes  $R_p \equiv \sin^2\theta_{13}/\alpha \approx 0.711$.  The CP violating phase $\delta_{cp}$ is not known at all and will one of the major interests at future neutrino facilities.

Parametrizing the interaction strength ${\hat A}$, Eq.~(\ref{hatAdef}), we find 
\beq
\label{ahatrho}
\hat A &=& \pm 6.50  ~10^{-2} R ~E[{\rm GeV}] \rho[{\rm gm/cm}^3] ~,
\eeq
with E[GeV] being the neutrino beam energy $E$ (in GeV) and $\rho$[gm/cm$^3$], is the average total density (in ${\rm gm/cm}^3$) of matter  through which the neutrino beam passes on its way to the detector (the matter having average proton-nucleon ratio $R$). For our calculations we are interested in experiments close to the earth's surface, so we take 
\beq
\rho[{\rm gm/cm}^3] &=& \rho_0 \nonumber \\
&=& 3 ~,
\eeq
the approximate mean density of the earth's mantle.

In the SNM, $\Delta_{L}$, defined in Eq.~(\ref{Deldeff}), may be parametrized in the high-energy limit as
\beq
\label{Deldefn}
\Delta_L &\approx&  3.05 \times 10^{-3} \frac { L[{\rm Km}] }{E[{\rm GeV}]} ~.
\eeq 
Here $ L[{\rm Km}] $ is  the baseline and $E[{\rm GeV}]$ is the neutrino beam energy.


\section{ Eigenvalue Expansions in the SNM}

The fact that $\alpha$ and $\sin\theta_{13}$ are naturally small in the SNM commonly motivates approximation schemes~\cite{f,ahlo,JO,Cer} based on first-order Taylor series expansions in one of these  small parameters, $\xi_i'$ (where $\xi_i'$ stands for $\alpha$ or $\sin^2{\theta_{13}})$.  For example, in Refs.~\cite{ahlo,JO} the oscillation probability is expanded in $\sin^2\theta_{13}$. Reference~\cite{f} makes use of an  expansion in the small parameter $\alpha$.

Although these expansions may be used effectively to simplify the theory, they come at a price~\cite{f}.  This price is that neither expansion gives accurate representations for all values of the interaction strength ${\hat A}$, including values in some regions of critical importance.

In our subsequent work~\cite{jhk2}, we will consider simplifying the  oscillation probability using the same expansions, but they will be used somewhat differently, in two stages. In the first stage, the eigenvalues are expanded, as below.  In the second, the expansions  will be used to simplify the oscillation probability.
In this work, the eigenvalue expansion will, of course be made before introducing $R_p$. For the expansion of the oscillation probabilities, $R_p$ may be introduced before the expansion is made because of its simpler analytic structure.

\subsection {Analytic Structure of Eigenvalues} 

The key for identifying which of the $\xi_i'$-expansions might be appropriate over specifics ranges of ${\hat A}$ is revealed by the analytic structure of the eigenvalue $ { \hat {\bar E}}_{\ell}(\xi)$. It is particularly important to identify the locations of its branch points when $\xi_i'=0$. Branch points identify where  a series expansion of $ { \hat {\bar E}}_{\ell}(\xi)$, or a function of it such as $<F(H_\nu)>$, would not converge.

We have seen that $ { \hat {\bar E}}_{\ell}(\xi)$ depends on $\xi$  entirely through the two functions $(U(\xi),V(\xi))$,
\beq
{ \hat {\bar E}}_{\ell}  (\xi)  &=& {\hat {\bar E}}_{\ell }(U(\xi),V(\xi)) ~,
\eeq
where 
\beq
U(\xi ) &\equiv& Re[d^{1/3}] \nonumber \\
V(\xi ) &\equiv& Im[d^{1/3}] ~,
\eeq
with $d$ defined in Eq.~(\ref{dalt}).

Recalling that $4\gamma^3 - \psi^2 > 0$, we conveniently write
\beq
U(\xi) &=& \frac{1}{2}  (\psi + i \sqrt{4\gamma^3-\psi^2 })^{1/3}  \nonumber \\
&+& (\psi -i \sqrt{4\gamma^3-\psi^2 })^{1/3}   ~,
\eeq
and
\beq
V(\xi) &=& \frac{1}{2i}  (\psi + i \sqrt{4\gamma^3-\psi^2 })^{1/3}  \nonumber \\
&-& (\psi -i \sqrt{4\gamma^3-\psi^2 })^{1/3}   ~.
\eeq
Branch points $\xi^B$ clearly occur for parameter values satisfying
\beq
\label{BPdef}
\psi^2(\xi^B) &=& 4 \gamma^3(\xi^B) ~.
\eeq
The locations of the branch points $\xi^B$ of $ {\hat {\bar E}}_{\ell}(\xi)$ are found as follows for the two small-parameter choices $\xi'_i$. 

When $\xi'_i=\alpha$, branch points as a function of the other parameters of the Standard Neutrino Model are found from Eq.~(\ref{BPdef}) with $\psi(\xi^B) $ and $\gamma(\xi^B) $ evaluated in terms of 
\beq
a(\xi^{B} ) &=& -(1+ {\hat A} ) \nonumber \\
b(\xi^{B}) &=& \hat A \cos^2\theta_{13} \nonumber \\
c(\xi^{B}) &=& 0 ~, 
\eeq
where $\xi^{B} = \xi|_{\alpha = 0}$.
The only real solution of Eq.~(\ref{BPdef}) is
\beq
{\hat A} &=& 0 ~.
\eeq

When $\xi'_i= \sin^2\theta_{13}$, branch points as a function of the other parameters of the SNM are found from Eq.~(\ref{BPdef}) with $\psi(\xi^B) $ and $\gamma(\xi^B) $ evaluated in terms of 
\beq
a(\xi^{B}) &=& -(1+\hat A + \alpha) \nonumber \\
b(\xi^{B}) &=& \alpha + \hat A \alpha \cos\theta_{12} + \hat A   \nonumber \\
c(\xi^{B}) &=& -\hat A \alpha \cos^2\theta_{12} ~,
\eeq 
where $\xi^{B} = \xi|_{\theta_{13} = 0}$.
The only real solution of Eq.~(\ref{BPdef}) is
\beq
\label{bpAtm}
{\hat A} &=& {\hat A}_0 \equiv  \frac{1-\alpha }{1 - \alpha \cos^2\theta_{12}} \nonumber \\
&\approx&  1 ~.
\eeq

\subsection{ The First-order Taylor Expansions}

Defining $\hat {\bar E}^{\xi}_\ell $ to be the first two terms of the Taylor series for $\hat {\bar E}_\ell $ expanded about $\xi=0$, we find 
\beq
\label{tayexp}
\hat {\bar E}^{\xi}_\ell &\equiv&  \hat {\bar E}_\ell|_{\xi=0}   + \xi \frac{\partial  \hat {\bar E}_\ell }{ \partial \xi} |_{\xi=0}   ~,
\eeq
where  $ \hat {\bar E}_\ell $ and $\partial \hat  {\bar E}_\ell /\partial \xi $ are easily obtained from Eqs.~(\ref{eqbarhatE1},\ref{abcvals}) and their derivatives. The results for $\xi = \sin^2\theta_{13} $ and $\xi = \alpha $ found in this way are given immediately below.

\subsubsection{ $\xi = \sin^2\theta_{13}$ }

Applying Eq.~(\ref{tayexp}) we find for $\xi = \sin^2\theta_{13}$ and below the corresponding branch point,
\beq
\label{evdiff1X}
{\hat {\bar E^{\theta } _{1} }} &=& \frac{1}{2 } ({\hat A} +\alpha  - \hat C_\theta ) \nonumber \\ 
&+& \frac{  \alpha R_p {\hat A} } { 2(1-y) \hat C_\theta } (2 -{\hat A}_0  - \alpha - \hat C_\theta )  \nonumber \\
&+&  \frac{ y \alpha R_p } { 2 \hat C_\theta } (2 - 4{\hat A}_0 + {\hat A}_0^2 -2 \alpha + 2 {\hat A}_0 \alpha) \nonumber \\
{\hat {\bar E^{\theta } _{2} }}
&=& \frac{1}{2 } ({\hat A} +\alpha  + \hat C_\theta ) \nonumber \\ 
&-& \frac{  \alpha R_p {\hat A} } { 2(1-y) \hat C_\theta } (2 -{\hat A}_0  - \alpha + \hat C_\theta )  \nonumber \\
&-&  \frac{ y \alpha R_p } { 2 \hat C_\theta } (2 - 4{\hat A}_0 + {\hat A}_0^2 -2 \alpha + 2 {\hat A}_0 \alpha)
\nonumber \\
{\hat {\bar E^{\theta}_{3} }}
&=& 1 +  \frac{ \alpha R_p \hat A }{1-y}  ~,
\eeq
where $ y \equiv \hat A / \hat A_0$, with ${\hat A}_0 \equiv {\hat A}^\theta_0$ the location of the branch point for the  $\sin^2\theta_{13} $ expansion and 
\beq
\hat C_\theta &=&  \sqrt{  {\hat A}^2 +\alpha^2 -2 {\hat A}  \alpha \cos{ 2 \theta_{12} } } ~.
\eeq
Note that ${\hat {\bar E^{\theta } _{2} }}
= {\hat {\bar E^\theta_{1} }}|_{ {\hat C}_\theta \to -{\hat C}_\theta }$. 
Above the branch point, $y>1$, ${\hat {\bar E^{\theta}_{3} }}$ and ${\hat {\bar E^{\theta}_{2} }}$ exchange roles. 

The lack of convergence at the branch point is manifest here by inspection, ${\it i.e. }$, by the appearance of simple pole at $y=1$. Although Eq.~(\ref{evdiff1X}) would suggest a pole in all three  eigenvalues,  ${\hat {\bar E^{\theta}_{1} }}$ is rather accurate for all values of $y$. The absence of a singularity in ${\hat {\bar E^{\theta}_{1} }}$ can be confirmed by a simple calculation that shows the coefficient of $(1-y)^{-1}$ vanishes at $y=1$.

\subsubsection{$\xi = \alpha $}

Taking  $R_p = \sin^2\theta_{13}/\alpha$, we may write $\cos^2\theta_{13} =  1- \alpha R_p$ and $\cos 2\theta_{13} =  1- 2 \alpha R_p$.  Then, for $\xi = \alpha $ and $y>0$ (above its corresponding branch point), 
\beq
\label{evdiff1Y}
{\hat {\bar E^\alpha_1 }} &=& \alpha \cos\theta_{12}^2 \nonumber \\
{\hat {\bar E^\alpha_{2} }} &=& \frac{1}{2} ( 1 + {\hat A}  - {\hat C}_\alpha ) \nonumber \\
&+& \alpha \frac{ \sin\theta_{12}^2} {2 } ( 1+ \frac{ 1 -  {\hat A} (1- 2 \alpha R_p)} { {\hat C}_\alpha } ) \nonumber \\
{\hat {\bar E^\alpha_{3} }} &=& \frac{1}{2 } (  1 +  {\hat A}  + {\hat C}_\alpha ) \nonumber \\
&+& \alpha \frac{ \sin\theta_{12}^2} {2 } ( 1- \frac{ 1 -  {\hat A} (1- 2 \alpha R_p) } { {\hat C}_\alpha } )~,
\eeq
where 
\beq
\hat C_\alpha &=& \sqrt{ (1 -  {\hat A })^2 + 4 \alpha R_p \hat A } ~.
\eeq
Note that  ${\hat {\bar E^\alpha_{2} }}   = {\hat {\bar E^\alpha_{3} }}|_{ {\hat C}_\alpha \to -{\hat C}_\alpha }$.  These expressions are identical to Eqs.~(18,19) of Freund~\cite{f}. 

In contrast to $\xi = \sin^2\theta_{13}$, the lack of convergence of the $\alpha$ expansion is not obvious from a casual examination of Eq.~(\ref{evdiff1Y}). 
Numerical comparison to the exact result confirms that ${\hat {\bar E^{\alpha}_{1} }}$ and ${\hat {\bar E^{\alpha}_{2} }}
$ are poor representations of the corresponding exact results in the vicinity of the branch point at $y = 0$. However, no evidence of the branch point is apparent in ${\hat {\bar E^{\alpha}_{3}}}$, which is rather accurate for all values of $y$.  Across the branch point at $y=0$, ${\hat {\bar E^{\alpha}_{1} }}$ and ${\hat {\bar E^{\alpha}_{2} }}$ exchange roles.

\subsection { Branch Points and Resonances }

Resonances are heralded by the appearance of minima in the EV differences. 
Two well-known neutrino resonances occur in the SNM. One of these, the ${\it solar}$ resonance, is found for relatively weak interaction strengths $\hat A \approx \alpha$. The other, the {\it atmospheric } resonance, occurs for stronger interactions, ${\hat A} \approx \cos{2\theta_{13}}$. 
The solar resonance occurs very close to the branch point identified with the $\alpha$ expansion, and the atmospheric resonance very close to the branch point identified with the $\sin^2\theta_{13}$ expansion. 

Using Eq.~(\ref{ahatrho}), the solar resonance appears for neutrinos of energy
\beq
E &=& E^{sol}[{\rm GeV}] \approx 0.325
\eeq
for underground experiments in the earth's mantle  ($ \rho[{\rm gm/cm}^3]  = \rho_0 \approx 3$).  We find, similarly, that for neutrinos of energy $E$, the solar resonance occurs in matter of density
\beq
\rho^{sol}[{\rm gm/cm}^3]  \approx E[{\rm GeV}]^{-1} ~,
\eeq
taking $\alpha=0.0317$ from the SNM.

Likewise, the
atmospheric resonance is found for neutrinos of  energy 
\beq
E &=& E^{atm}[{\rm GeV}] \approx 9.79 ~, 
\eeq
also for underground experiments in the earth's mantle.  At energy $E$, it occurs at a density of
\beq
\rho^{atm}[{\rm gm/cm}^3]  \approx 29.4  / E[{\rm GeV}] ~,
\eeq
taking  $ \cos{2 \theta_{13} } = 0.955$ from the SNM.

Because of the close correlation between the branch points and the resonances, there is also a close correlation between resonances and viable approximation schemes. Note, however, that the branch point in the $\alpha$ expansion near the solar resonance affects both neutrino and anti-neutrino scattering since its actual location is $\alpha =0$.  

\subsection{Discussion}

We have seen in this section that expansions of the eigenvalues  in a small parameter $\xi_i'$ of the SNM (where $\xi_i'$ stands for $\alpha$ or $\sin^2{\theta_{13}})$ must be made carefully, since the EV are not analytic everywhere.  

Nevertheless, as noted, first-order Taylor series representations of the EV are commonly used to simplify the theory.  Because these expansions do not  give an  accurate representation the EV for all values of the interaction strength ${\hat A}$, it is important to identify the regions where the theoretical errors of the expanded EV might be acceptable and lead to accurate representations.  We address this in the next section.

\section{ Approximating $\mathcal{P}(\nu_i  \rightarrow \nu_f) $ with expanded EV} 

In this section, we begin our assessment of common procedures used to simplify calculation of the oscillation probability by expanding it  in one of the small parameters of the SNM.  Freund observed~\cite{f} that the $\alpha$-expansion, although useful, could not be used near the solar resonance where ${\hat A}=\alpha$.  However, no understanding of the limitations of the $\sin^2\theta_{13}$-expansion appears in the literature.

We have shown above that the applicability of both expansions is limited by the presence of branch points in the analytical structure of the eigenvalues. The branch point responsible for the failure of the $\alpha$-expansion is located at $\hat A = 0$ when $\alpha=0$, and the branch point responsible for the failure of the $\sin^2\theta_{13}$ expansion is located at $\hat A = \hat A_0$ when $\sin^2\theta_{13}=0$, where $\hat A_0$ is defined in Eq.~(\ref{bpAtm}).

We make our assessment numerically, comparing the oscillation probability calculated from Eqs.~(\ref{POPLdef1b},\ref{a:ppsin1c2})  using the exact eigenvalues to that calculated from Eqs.~(\ref{POPLdef1b},\ref{a:ppsin1c2}) using eigenvalues expanded in one of the small parameters of the SNM. 


\subsection { Assessing Oscillation Probabilities Expressed in terms of $\xi$-expanded Eigenvalues }

Theoretical errors characterizing an approximation scheme may emerge numerically only from an examination of the dependence of $P^{e\mu}(\Delta_L,{\hat A})$ on $\Delta_L$ and $\hat A$. In this section we discuss how we will do this.

It is convenient to discuss the oscillation pattern in terms of the location of the first maxima of the functions $ \Delta_L j_0 ({\hat \Delta}_\ell)$ appearing in the expressions for the partial oscillation probabilities. These peaks occur at $\Delta_L =  \Delta_L^{(\ell)}$, with    
\beq
\label{L0defa}
\Delta_L^{(\ell)}  &=& \frac{\pi} {2 \Delta {\hat{\bar E}} [\ell] } 
\eeq
closely related to the period $ P_\ell $ of $ j_0 ({\hat \Delta}_\ell)$, 
\beq
P_\ell &\equiv& 4 \Delta_L^{(\ell)} = \frac{2\pi}{ \Delta {\hat{\bar E}}[\ell] } ~.
\eeq

Because the exact eigenvalues never cross, the ordering of $ {\hat{\bar E}}_\ell$ is the same as it is in the vacuum, namely $ {\hat{\bar E}}_3 > {\hat{\bar E}}_2  > {\hat{\bar E}}_1 $. It can inferred from this that all $\Delta {\hat {\bar E}}[\ell]$ remain positive, and, in addition, that
\beq
\label{orderDE}
\Delta {\hat {\bar E}}[2] &>& \Delta {\hat {\bar E}}[3] >0 \nonumber \\
\Delta {\hat {\bar E}}[2] &>& \Delta {\hat {\bar E}}[1] > 0 ~.
\eeq
We see, in general, that for small ${\hat A}$, $\Delta {\hat {\bar E}}[1] > \Delta {\hat {\bar E}}[3]$, and, for large ${\hat A}$, $\Delta {\hat {\bar E}}[2] > \Delta {\hat {\bar E}}[3]$. 

It is also clear that $\Delta {\hat {\bar E}}[2] $ is always the largest eigenvalue difference. Consequently,  $P_2$ is always the smallest of the three periods, thus characterizing the most rapidly varying Bessel function. The relative sizes of $P_\ell$ are easily worked out in specific cases. In the vacuum, 
\beq
\label{Pvac}
P_1 &=&  2 \pi/(1-\alpha) \nonumber \\
P_2 &=&  2 \pi  \nonumber \\
P_3 &=& 2 \pi / \alpha ~,
\eeq 
evaluated from differences of the vacuum eigenvalues  appearing  in Eq.~(\ref{h0st2}).

In the SNM, we find that 
\beq
\label{EVdifSNM}
P_3 &>& P_1 ~\rm{for}~ \hat A < \hat A_{2} \nonumber \\
P_1 &>& P_3 ~\rm{for}~  \hat A > \hat A_{2}~, 
\eeq
where
\beq
\label{hatA2}
\hat A_{2} &=& 0.538  
\eeq 
is the value of $\hat A$ at which $P_3 = P_1$ ($\Delta {\hat {\bar E}}[1] = \Delta {\hat {\bar E}}[3]$). 
 
With $\Delta {\hat {\bar E}}[2]$ the largest eigenvalue difference, $P_2$ is always the smallest of the three periods.  Thus, the value of $\Delta_L$ at the first peak of $\Delta_L j_0({\hat \Delta}_2)$ is a natural scale. 

\subsection {Regions of Maximum Sensitivity }

Sensitivity to the approximation scheme should be most manifest within regions where all three Bessel functions of $P^{ab}( \Delta_L,{\hat A})$ are 
of similar size and interfere.  This will happen once $\Delta_L$ becomes comparable to the first peak of its most slowly varying $ \Delta_L j_0 ({\hat \Delta}_\ell)$, which occurs at either $\Delta_L = \Delta_L^{(3)}$ or $\Delta_L = \Delta_L^{(1)}$. In general, the sensitivity to approximations increases as the distance $\Delta_L$ to the most distant peak increases.

In the vacuum, the first peak of the most slowly varying $ \Delta_L j_0 ({\hat \Delta}_\ell)$ is  always at $\Delta_L  = \Delta_L^{(3)}$, readily established by the vacuum eigenvalue differences given in Eq.~(\ref{h0st2}). 
The corresponding baseline is obtained from Eqs.~(\ref{Deldeff},\ref{L0defa}), $L^{(3)}[{\rm Km}] \approx 5170  \pi E[{\rm GeV}]$.

In the SNM, the most slowly varying Bessel function, established from Eqs.~(\ref{EVdifSNM},\ref{hatA2}), is  $ \Delta_L j_0 ({\hat \Delta}_3)$ when  ${\hat A}< \hat A_{2}$ and $ \Delta_L j_0 ({\hat \Delta}_1)$
for ${\hat A}> \hat A_{2}$.

\subsubsection {Regions of Maximum Sensitivity for Fixed $\hat A$ }

Consider first the variation of $P^{(ab)}(\Delta_L, \hat A)$ with $\Delta_L$ for a given value of $\hat A$ in the SNM. 

According to Eq.~(\ref{EVdifSNM}), for $\hat A > \hat A_2$, $ \Delta_L j_0 ({\hat \Delta}_1)$ is the most slowly varying (having the larger period), and the region of maximum sensitivity is
\beq
\label{fixedAG}
\Delta_L &>& \Delta_L^{(1)} \equiv \frac{\pi} {2 \Delta {\hat{\bar E}} [1] } \eeq
with $\Delta {\hat{\bar E}} [1]$ evaluated at $\hat A$.

Similarly, for $\hat A < \hat A_2$,  $ \Delta_L j_0 ({\hat \Delta}_3)$ is the most slowly varying.  According to Eq.~(\ref{L0defa}), its peak occurs where $ \Delta_L = \pi/(2 \Delta {\hat{\bar E}} [3] )$. Thus, the region of maximum sensitivity is
\beq
\label{fixedAL}
\Delta_L &>& \Delta_L^{(3)} \equiv \frac{\pi} {2 \Delta {\hat{\bar E}} [3] }  ~. 
\eeq
with $\Delta {\hat{\bar E}} [3]$ again evaluated at $\hat A$.

\subsubsection {Regions of Maximum Sensitivity for Fixed $\Delta_L$ }

Consider next the variation of $P^{(ab)}(\Delta^0_L, \hat A)$ with $\hat A$ for given $\Delta^0_L$ in the SNM. 

According to Eq.~(\ref{EVdifSNM}), for $\hat A > \hat A_2 $, $ \Delta_L j_0 ({\hat \Delta}_1)$ is the most slowly varying. According to Eq.~(\ref{L0defa}), its peak occurs where $ \Delta^0_L = \pi/(2 \Delta {\hat{\bar E}} [1])$. Thus, the region of maximum sensitivity is
\beq
\label{fixedDG}
\Delta {\hat{\bar E}} [1]  &>& \frac {\pi }{2 \Delta^0_L } 
\eeq
with $\Delta {\hat{\bar E}} [1]$ evaluated at $\hat A$. 

Similarly, for $\hat A < \hat A_2 $, we find from  Eq.~(\ref{fixedAG}) that $ \Delta_L j_0 ({\hat \Delta}_3)$ is the most slowly varying, and the region of maximum sensitivity is 
\beq
\label{fixedDL}
\Delta {\hat{\bar E}} [3]  &>& \frac {\pi }{2 \Delta^0_L } ~.
\eeq

\section{ Numerical Study of $\mathcal{P}(\nu_e  \rightarrow \nu_\mu)$ with expanded EV}

Our main interest in the present section is to map out the regions where each of the small-parameter expansions is capable of simplifying $P^{ab}( \Delta_L,{\hat A})$. We do this by comparing three calculations using $\mathcal{P}(\nu_e  \rightarrow \nu_\mu)$ taken from our exact Hamiltonian formulation.  

One of these is a calculation of the exact oscillation probability obtained in our Hamiltonian formulation.  For this we use the expressions in Eqs.~(\ref{POPLdef1a},\ref{POPLdef1b}) evaluated with the exact EV. The other two are calculations of our $\xi$-expanded oscillation probability for each of the small parameters of the SNM.  For these we evaluate Eqs.~(\ref{POPLdef1a},\ref{POPLdef1b}) using  the $\xi$-expanded EV.  
For $\xi=\alpha$, we use the $\alpha$-expanded EV given in Eq.~(\ref{evdiff1Y}), and for $\xi=\sin^2\theta_{13}$ we use the $\sin^2\theta_{13}$-expanded EV given in Eq.~(\ref{evdiff1X}).  The calculation with the $\xi$-expanded EV would, of course, coincide with the exact calculation in the vacuum. Differences therefore reflect medium effects.

The extent to which our oscillation probability evaluated with one of the $\xi$-expanded EV agrees with the exact result indicates regions in which it may be possible to obtain, at least in principle, a simple $\xi$-expanded expression for the oscillation probability in good agreement with the exact result.  In a subsequent paper~\cite{jhk1}, we make a similar comparison between the exact oscillation probability and the approximate ones given in Refs.\cite{ahlo,f}. 

From the numerical results we obtain in Ref.~\cite{jhk1} and the present paper, we will be able to  identify regions in which both (1) the exact oscillation probability and the results of Refs.\cite{f,ahlo} are in poor agreement; and, (2) the exact oscillation probability and the $\xi$-expanded result are in excellent agreement.  The regions where both of these conditions are satisfied indicate where it might be possible to improve the results found in Refs.\cite{f,ahlo} using our Hamiltonian formulation. We explore this possibility in yet another paper~\cite{jhk2}.

Equations~(\ref{Deldeff},\ref{ahatrho}) provide a means to extrapolate the results in any figure to a variety of baseline values, medium properties, and neutrino energies within the particular regions shown in that figure. In  the earth's mantel, where the average $Z/N=1/2$, Eqs.~(\ref{ahatrho},\ref{Deldeff}) become
\beq
\label{Deldefalt3}
 L[{\rm Km}] &=& 1.08\times 10^4 \frac{ {\hat A} \Delta_L }{ \rho[{\rm gm/cm}^3] } \nonumber \\
E[{\rm GeV}] &=& 30.8 \frac{ {\hat A}}{ \rho[{\rm gm/cm}^3] } ~.
\eeq

\subsection { $\Delta_L$-Dependence of $\mathcal{P}(\nu_e  \rightarrow \nu_\mu)$ }  

We begin our exploration of the extent to which a particular $\xi$-expansion is capable of simplifying the oscillation probability by examining $P^{(e\mu)}(\Delta_L, \hat A)$ vs  $\Delta_L$ for particular values of $\hat A$. One value of $\hat A$ is chosen near the solar resonance and another near the atmospheric resonance.  For each choice of $\xi$ and $\hat A$, we compare the exact result to Eq.~(\ref{tayexp}).  

We first examine $\mathcal{P}(\nu_e  \rightarrow \nu_\mu)$ below the solar resonance, at ${\hat A} = 0.0102$. For this value of $\hat A$, $\Delta {\hat{\bar E}} [3]  \approx 0.0294$, Eq.~(\ref{fixedAL}) specifies that the approximate oscillation probability becomes  sensitive to approximations for $\Delta_L > \Delta_L^{(3)} \approx 53$.

Our calculations for ${\hat A} = 0.0102$ are shown in Fig.~\ref{PL12SolD}.
We see from this figure that the $\alpha$-expanded oscillation probability 
begins to departs from the exact result at large $\Delta_L$ meaning, as expected, that  the   $\alpha$ expansion breaks down in the vicinity of the solar resonance. The sensitivity to medium effects shows up already for $\Delta_L \approx 20$, which is smaller than  $\Delta_L^{(3)} \approx 53 $ estimated using Eq.~(\ref{fixedAL}).  On the other hand,  Eq.~(\ref{tayexp}) evaluated with eigenvalues expanded to first order in $\sin^2\theta_{13}$ agree well with the exact result at large $\Delta_L$ showing that the $\sin^2\theta_{13}$-expanded oscillation probability is capable of providing an excellent approximation in the vicinity of the solar resonance.

\begin{figure}
\centerline{\epsfig{file=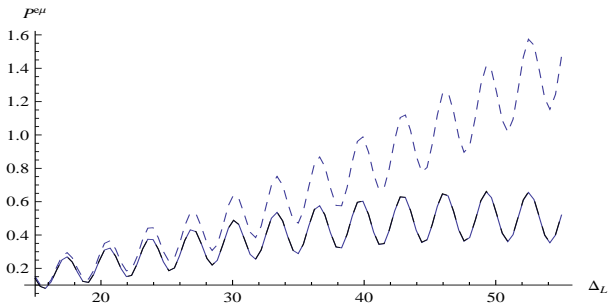,height=4cm,width=8.cm}}
\caption{  $P^{e\mu}(\Delta_L,{\hat A})$ in the solar-resonance region (${\hat A}=0.0102$) over the interval $15<\Delta_L<55$ for neutrinos in matter. Parameters are taken from the SNM. Exact result (solid curve).  Our expression for the  oscillation probability evaluated with the $\alpha$-expanded EV (medium-dashed curve). Our expression for the   oscillation probability evaluated with the $\sin^2\theta_{13}$-expanded EV (long-dashed curve) .}
\label{PL12SolD}
\end{figure}

For this small value of ${\hat A}$, we find that position of the first peak of the exact oscillation probability, $\Delta_L \approx 1.58$, coincides almost exactly with the location of the peak of the most rapidly varying Bessel function, $j_0({\hat \Delta}_2)$.
From Eq.~(\ref{Deldefalt3}), we note  that the oscillation probability at $\Delta_L=20$ in Fig.~\ref{PL12SolD}, where the approximate calculation begins to break down in the solar resonance region with the $\alpha$-expanded EV, would correspond to a measurement of $105$ MeV neutrinos propagating in the earth's mantel at a baseline $734$ Km.  

We next examine $\mathcal{P}(\nu_e  \rightarrow \nu_\mu)$ for ${\hat A} = 0.8$, a value of ${\hat A}$ near the atmospheric resonance. Taking the exact eigenvalue difference $\Delta {\hat{\bar E}} [1]  \approx 0.328$ at ${\hat A} = 0.8$, Eq.~(\ref{fixedAG}) specifies that the desired sensitivity of the oscillation probability to approximations should become apparent at $\Delta_L \approx 4.8$.  Our calculations for ${\hat A} = 0.8$ are shown in Fig.~\ref{PL12AtmD}.

The exact result shown in Fig.~\ref{PL12AtmD} (solid curve) begins to differ from the $\sin\theta_{13}$-expanded result (long-dashed curve) at  $ \Delta_L \approx 2$, which occurs  a bit before $ \Delta_L  \approx 4.8$, where all three Bessel functions fully contribute. Because the long-dashed curve begins to depart from the solid curve at large $\Delta_L$, these results confirm that $P^{e\mu}(\Delta_L,{\hat A})$ evaluated with  $\sin^2\theta_{13}$-expanded EV  breaks down near the atmospheric resonance with medium effects included. The failure of the $\sin\theta_{13}$-expansion becomes increasingly apparent as $\Delta_L$ increases to larger  $\Delta_L$.

The fact that the exact result (solid curve) and $\alpha$-expanded result (medium-dashed curve) seem to completely overlap demonstrates that $P^{e\mu}(\Delta_L,{\hat A})$ evaluated with the $\alpha$-expanded EV is capable of becoming a completely acceptable approximation near the atmospheric resonance. 

\begin{figure}
\centerline{\epsfig{file=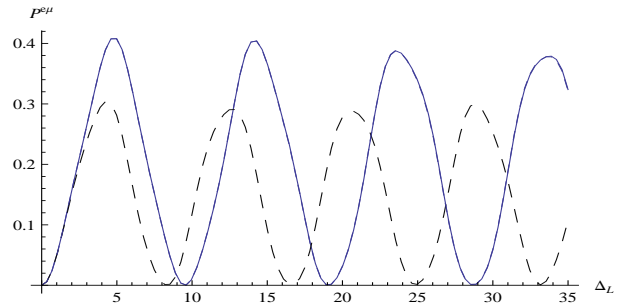,height=4cm,width=8.cm}}
\caption{  $P^{e\mu}(\Delta_L,{\hat A})$ in the atmospheric resonance region (${\hat A}=0.8$) over the interval $0<\Delta_L<35$ for neutrinos in matter. Parameters are taken from the SNM. Exact result (solid curve).  Our expression for the   oscillation probability evaluated with the $\alpha$-expanded EV (medium-dashed curve).}
\label{PL12AtmD}

\end{figure}
For this larger value of ${\hat A}$, the first peak of the oscillation probability, at $\Delta_L \approx 1.58 $, nearly coincides with the peak of the most rapidly varying Bessel function, $j_0({\hat \Delta}_2)$, at $\Delta_L \approx 1.5$. 
Applying Eq.~(\ref{Deldefalt3}),  we see that the oscillation probability at the value of  $\Delta_L$ where medium effects begin to become apparent in Fig.~\ref{PL12AtmD} would correspond to neutrinos of energy $E[{\rm GeV}] \approx 4.5$ propagating in matter of density similar to the average density of the entire earth, $\rho[{\rm gm/cm}^3]  = 5.52$ at a baseline of $7830$ Km (for comparison, the average earth radius is $6370$ Km).

\subsection{${\hat A}$-Dependence of $\mathcal{P}(\nu_e  \rightarrow \nu_\mu)$ }

We next compare oscillation probabilities over various ranges of $\hat A$.
For a given range of $\hat A$, whether or not the three Bessel functions maximally interfere depends on the choice of $\Delta_L$, which is determined by Eqs.~(\ref{fixedDG},\ref{fixedDL}) depending on whether $\hat A > \hat A_2$ or $\hat A < \hat A_2$, respectively. 

\subsubsection{ $0< \hat A < 0.2$ }

Numerical studies using Eqs.~(\ref{fixedDG},\ref{fixedDL}) show that for $0 <{\hat A} < 0.2 $ taking $\Delta_L=60$ is sufficient to ensure that the three Bessel functions maximally interfere. In Fig.~\ref{PL12ASol}, we compare the oscillation probabilities in this region.  These results confirm that the expansion in $\sin^2\theta_{13}$ is a reasonably good approximation within the solar resonance region, whereas the expansion in $\alpha$ is evidently not.

\begin{figure}
\centerline{\epsfig{file=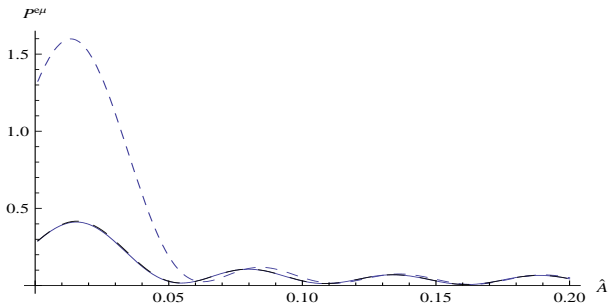,height=4cm,width=8.cm}}
\caption{  $P^{e\mu}(\Delta_L,{\hat A})$ for $\Delta_L=60$ over the interval $0<{\hat A}<0.2$, for neutrinos in matter. Parameters are taken from the SNM. Exact result (solid curve).  Our expression for the   oscillation probability evaluated with the $\alpha$-expanded EV (medium-dashed curve). Our expression for the   oscillation probability evaluated with the $\sin^2\theta_{13}$-expanded EV (long-dashed curve) .}
\label{PL12ASol}
\end{figure}

\subsubsection{ $0.2 < \hat A < \hat A_2$ }

Numerical studies using Eqs.~(\ref{fixedDG},\ref{fixedDL}) show that for $0 <{\hat A} < \hat A_2 $ taking $\Delta_L=10$ is sufficient to ensure that the three Bessel functions maximally interfere. In Fig.~\ref{PL12ATrn}, we compare the oscillation probabilities in this region.  These results confirm that both the expansion in $\sin^2\theta_{13}$ and the expansion in $\alpha$ are reasonably good approximations here.

\begin{figure}
\centerline{\epsfig{file=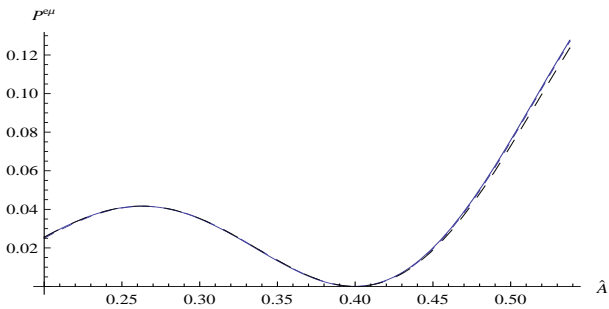,height=4cm,width=8.cm}}
\caption{  $P^{e\mu}(\Delta_L,{\hat A})$ for $\Delta_L=10$ over the interval $0.2<{\hat A}< \hat A_2$, for neutrinos in matter. Parameters are taken from the SNM. Exact result (solid curve).  Our expression for the   oscillation probability evaluated with the $\alpha$-expanded EV (medium-dashed curve). Our expression for the   oscillation probability evaluated with the $\sin^2\theta_{13}$-expanded EV (long-dashed curve) .}
\label{PL12ATrn}
\end{figure}
Results shown here apply over the same range of neutrino energy, baselines, and medium properties as those given in Eqs.~(\ref{Deldefalt3}). The extrapolation applies, of course, only within the region  $0.2 <{\hat A} < \hat A_2$. 

\subsubsection{ $\hat A_2 < \hat A < 0.8$ }

Numerical studies using Eqs.~(\ref{fixedDG},\ref{fixedDL}) show that for $\hat A_2  <{\hat A} < 0.8 $ taking $\Delta_L=4$ is sufficient to ensure that the three Bessel functions maximally interfere. In Fig.~\ref{PL12APreAtm}, we compare the oscillation probabilities over this region.  These results confirm that the expansion in $\alpha$ is a reasonably good approximation below the atmospheric resonance.  The onset of the failure of the $\sin^2\theta_{13}$-expanded EV near the atmospheric resonance begins to become visible for ${\hat A} > \hat A_2$.
\begin{figure}
\centerline{\epsfig{file=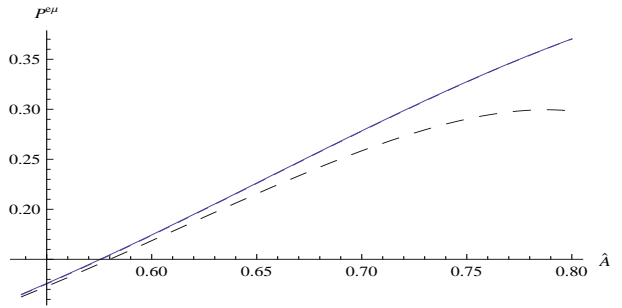,height=4cm,width=8.cm}}
\caption{  $P^{e\mu}(\Delta_L,{\hat A})$ for $\Delta_L=4$ over the interval $\hat A_2 <{\hat A}<0.8$, for neutrinos in matter. Parameters are taken from the SNM. Exact result (solid curve).  Our expression for the   oscillation probability evaluated with the $\alpha$-expanded EV (medium-dashed curve). Our expression for the   oscillation probability evaluated with the $\sin^2\theta_{13}$-expanded EV (long-dashed curve) .}
\label{PL12APreAtm}
\end{figure}
Results shown here apply over the same range of neutrino energy, baselines, and medium properties as those given in Eqs.~(\ref{Deldefalt3}). The extrapolation applies, of course, only for  $\hat A_2 <{\hat A} < 0.8$.

\subsubsection{ $\hat 0.8 < \hat A < 1.2 $ }

Numerical studies using Eqs.~(\ref{fixedDG},\ref{fixedDL}) show that for $\hat 0.8   <{\hat A} < 1.2 $ taking $\Delta_L=4$ is sufficient to ensure that the three Bessel functions maximally interfere. In Fig.~\ref{PL12AAtm}, we compare the oscillation probabilities over this region.  These results confirm that the expansion in $\alpha$ is a reasonably good approximation across the atmospheric resonance.  The $\sin^2\theta_{13}$-expanded oscillation probability is not shown because it fails here.
\begin{figure}
\centerline{\epsfig{file=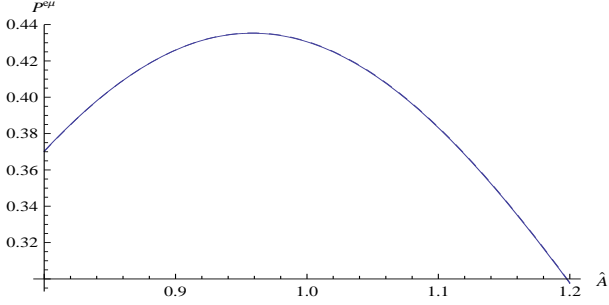,height=4cm,width=8.cm}}
\caption{  $P^{e\mu}(\Delta_L,{\hat A})$ for $\Delta_L=4$ over the interval $0.8  <{\hat A}<1.2$, for neutrinos in matter. Parameters are taken from the SNM. Exact result (solid curve).  Our expression for the   oscillation probability evaluated with the $\alpha$-expanded EV (medium-dashed curve). Our expression for the   oscillation probability evaluated with the $\sin^2\theta_{13}$-expanded EV (long-dashed curve) .}
\label{PL12AAtm}
\end{figure}
Results shown here apply over the same range of neutrino energy, baselines, and medium properties as those given in Eqs.~(\ref{Deldefalt3}). The extrapolation applies, of course, only for  $0.8 <{\hat A} < 1.2$.

\subsubsection{ $1.2 < \hat A < 2.5$ }

Numerical studies using Eqs.~(\ref{fixedDG}) show that for $1.2 < {\hat A} < 2.5$ taking $\Delta_L = 4$ is sufficient to ensure that the three Bessel functions maximally interfere.  In Fig.~\ref{PL12AG}, we compare the oscillation probabilities over this region.  These results show that $P^{e\mu}(\Delta_L,{\hat A})$ evaluated with the $\alpha$-expanded EV is relatively accurate here.  The $\sin^2\theta_{13}$-expanded oscillation probability agrees with the exact result reasonably well for $\hat A > 1.6$. 
\begin{figure}
\centerline{\epsfig{file=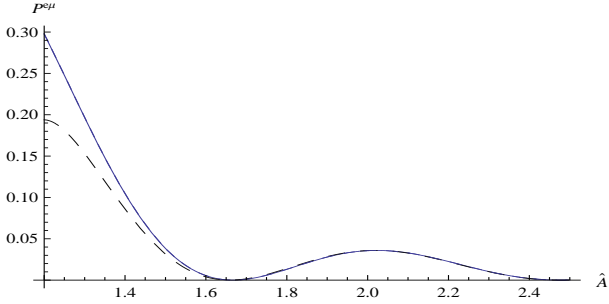,height=4cm,width=8.cm}}
\caption{  $P^{e\mu}(\Delta_L,{\hat A})$ {\it vs} ${\hat A}$ for neutrinos in matter over the interval  $1.2 < {\hat A} < 2.5$  taking $\Delta_L=4$. Parameters are taken from the SNM. Exact result (solid curve).  Our expression for the   oscillation probability evaluated with the $\alpha$-expanded EV (medium-dashed curve). Our expression for the   oscillation probability evaluated with the $\sin^2\theta_{13}$-expanded EV (long-dashed curve) .}
\label{PL12AG}
\end{figure}

From Eq.~(\ref{Deldefalt3}), we see that with the value $\Delta_L=4$, and in matter of mean density similar to that of the average density of the entire earth, $\rho[{\rm gm/cm}^3]  = 5.52$, the breakdown  of the $\sin^2\theta_{13}$-expanded EV starts to becomes less visible for ${\hat A} > 1.2$, corresponding to  neutrinos of energy $E[{\rm GeV}] \approx 6.2$ and baselines greater than $ L[{\rm Km}] = 23,500$ (considerably larger than the diameter of the earth).

\subsubsection{$-0.5 < \hat A < 0.5$ }

In Fig.~\ref{PL12Apm}, we compare the oscillation probability calculated with the exact eigenvalues to  the oscillation probability calculated with the expanded eigenvalues for $-0.5<{\hat A} < 0.5$, taking $\Delta_L=35$.
Figure~\ref{PL12Apm} confirms the earlier observations that the expansion in $\sin^2\theta_{13}$ is a valid approximation within the solar resonance region, whereas the expansion in $\alpha$ is not for both neutrinos and anti-neutrinos.  
\begin{figure}
\centerline{\epsfig{file=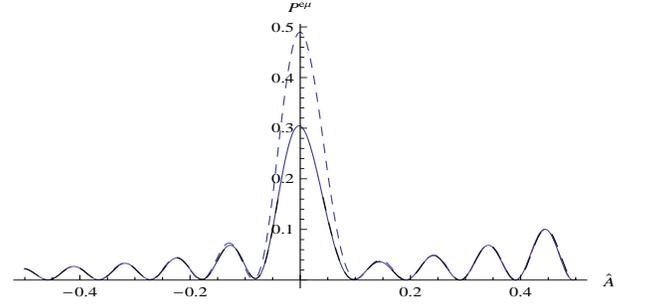,height=4cm,width=8.cm}}
\caption{  $P^{e\mu}(\Delta_L,{\hat A})$ for $\Delta_L=35$ over the interval $-0.5<{\hat A}<0.5$, for neutrinos and anti-neutrinos in matter. Parameters are taken from the SNM. Exact result (solid curve).  Our expression for the   oscillation probability evaluated with the $\alpha$-expanded EV (medium-dashed curve). Our expression for the   oscillation probability evaluated with the $\sin^2\theta_{13}$-expanded EV (long-dashed curve) .}
\label{PL12Apm}
\end{figure}

\subsection{ Intrinsic Limitations  }

It is important to note that when using expanded EV, the range of validity of the oscillation probability $P^{ab}( \Delta_L ,{\hat A})$ is limited even when it is evaluated using the convergent eigenvalue expansion.  This arises because the error in the phase of the trigonometric functions  $\sin(\Delta_L \Delta {\hat{\bar E}}^\xi[\ell])$ grows with $\Delta_L $. Eventually, with increasing $\Delta_L $, the error in the phase will lead to an unacceptable error in $P^{ab}( \Delta_L ,{\hat A})$.

When spurious effects of this type would show up from a comparison of the exact partial oscillation probability evaluated with the exact ${\hat{\bar E}}_\ell$ to the exact partial oscillation probability evaluated with the convergent expansion of ${\hat{\bar E}}_\ell$. A divergence between the calculated results of these two calculations herald the limit of $\Delta_L$ beyond which the use of expanded eigenvalues breaks down for $P^{ab}( \Delta_L ,{\hat A})$.

It is clear from Fig.~\ref{PL12SolD} that the region of validity for both the $\sin^2\theta_{13}$-expansion and the $\alpha$-expansion extend out as far as $\Delta_L = 35$.

\subsection{Discussion}

The results of this section suggest a natural division of the full range of $\hat A$ into regions. One significant region, which we call the ${\it solar }$ resonance region, covers the range $0 < \hat A < 0.2 $. The solar-resonance region contains, of course, the solar resonance at $\hat A = \alpha$. Another is the atmospheric resonance region containing the atmospheric resonance at $\hat A \approx  \cos 2\theta_{13}$. This region covers the interval $\hat A_2  < {\hat A} < 2.0$. 

The region between the solar and atmospheric regions, $0.2 < {\hat A}  < \hat A_2 $, is a ${\it transition}$ region in the sense that within it the expansion in $\alpha$ is improving rapidly and the expansion in $\sin^2 \theta_{13}$ rapidly deteriorating.  Within this region, the EV expansions in $\sin^2 \theta_{13}$ and $\alpha$ are of comparable accuracy. 
The ${\it asymptotic}$ region covers the interval ${\hat A}  > 2.0$, where the EV's are approaching their asymptotic behavior. 

In future work, we will make a comparison of the our oscillation probability evaluated with the exact eigenvalues to the approximate oscillation probability given in  Refs.~\cite{f,ahlo}. By comparing those results to our oscillation probability evaluated with $\xi$-expanded EV found in the present paper, we will be able to identify regions where our Hamiltonian formulation might lead to more effective approximations.  Based on this information, we will subsequently present any new approximate results we believe be helpful.

\section{ Summary and Conclusions}

We have presented exact, closed-form expressions for the neutrino oscillation probabilities  in matter using our Hamiltonian formulation within the framework of the Standard Neutrino Model assuming 3 Dirac Neutrinos. Our goal is to benchmark approximate formulations  having known difficulties arising from expansions commonly used to model neutrino and anti-neutrino experiments envisioned for future neutrino facilities.  

We have shown explicitly that for small $\alpha$ and $\sin\theta_{13}$ there are branch points in the analytic structure of the eigenvalues that lead to singular behavior of expansions near the solar and atmospheric resonance. The numerical calculations presented  indicate regions in which  the small-parameter expansions are guaranteed fail and should be avoided in practice.

\appendix


\section{ Time-evolution operator $S(T)$ in Our Hamiltonian Formulation}
\label{a:LG}

Our Hamiltonian formulation was presented in Sect.~\ref{LF}.  
In this appendix, and in accord with the discussion in Sect.~\ref{LF},  we give explicit analytic formulae for the time-evolution operator $ S(L) = e^{-i  {\hat {\bar H}}_\nu \Delta_L}  $ in terms of ${\hat {\bar E}}_\ell $ and the parameters defining ${\hat {\bar H}}_{\nu}$.

\subsection{ Expression for $S(T)$ in terms of $ F^{ab}_{\ell}  $ }
\label{A:Sab}

Introducing dimensionless variables, 
the time-evolution operator $ S(T)$ in Eq.~(\ref{lagrf}) may be written, 
\beq
\label{lagrf1}
S^{ab}(L) &=& \sum_{\ell}  F^{ab}_{\ell}  \cos{ 2{\hat{\bar E}}_{\ell} \Delta_L  } 
\nonumber \\
&-&  i \sum_{\ell}  F^{ab}_{\ell} \sin{ 2{\hat{\bar E}}_{\ell} \Delta_L  }   ~.  \eeq
Here $ F^{ab}_{\ell} $, defined in Eq.~(\ref{Wdef}), may be found by evaluating Eq.~(\ref{lagrf1}) at $L=0$ and obeys the normalization condition 
\beq
\label{lagrfN}
\delta_{ab}  &=& \sum_{\ell} F^{ab}_{\ell} ~.
\eeq

Using the identity $\cos{ 2 \beta }  \equiv 1- 2\sin^2{ \beta}$,  we write
\beq
\label{lagrf3}
S^{ab}(L) &=& \sum_{\ell}  
F^{ab}_{\ell}  ( 1 - 2 \sin^2{ {\hat{\bar E}}_{\ell} \Delta_L  } ) 
\nonumber \\
&-&  i \sum_{\ell} 
F^{ab}_{\ell} \sin{ 2{\hat{\bar E}}_{\ell} \Delta_L  }  \nonumber \\
&=& \sum_{\ell} F^{ab}_{\ell} 
- 2 \sum_{\ell}   F^{ab}_{\ell}  \sin^2{ {\hat{\bar E}}_{\ell} \Delta_L  } \nonumber \\
&-&  i \sum_{\ell} 
F^{ab}_{\ell} \sin{ 2{\hat{\bar E}}_{\ell} \Delta_L  }    ~.
\eeq
Now, using the normalization, Eq.~(\ref{lagrfN}),
\beq
\label{lagrf4}
S^{ab}(L) &=& \delta_{ab} -  2 \sum_{\ell}  
F^{ab}_{\ell}  \sin^2{ {\hat{\bar E}}_{\ell} \Delta_L  } \nonumber \\
&-&  i \sum_{\ell} 
F^{ab}_{\ell} \sin{ 2{\hat{\bar E}}_{\ell} \Delta_L  }    ~.
\eeq
Thus,  for calculating $ P^{ab}(L) $ in Eq.~(\ref{oscprob0}), 
\beq
\label{lagrf5}
Re[ S^{ab}(t',t) ] &=&  \delta_{ab} -  2 \sum_{\ell}  
Re[ F^{ab}_{\ell} ] \sin^2{ {\hat{\bar E}}_{\ell} \Delta_L  } \nonumber \\
&+& \sum_{\ell}  Im [F^{ab}_{\ell} ] \sin{ 2{\hat{\bar E}}_{\ell} \Delta_L  }    \nonumber \\
Im[ S^{ab}(t',t) ] &=&  -  2 \sum_{\ell}  
Im[ F^{ab}_{\ell} ] \sin^2{ {\hat{\bar E}}_{\ell} \Delta_L  } \nonumber \\
&-& \sum_{\ell}  Re[ F^{ab}_{\ell} ] \sin{ 2{\hat{\bar E}}_{\ell} \Delta_L  } ~.
\eeq

\subsection{ Analytic Expressions for ${\hat {\bar W}}_{0}[\ell]$, ${\hat {\bar W}}_{sin}[\ell]$, and ${\hat {\bar W}}_{cos}[\ell]$   in terms of $H_\nu$}
\label{Wvals}

In this appendix we give exact, analytic expressions the $S(T)$ in terms of ${\hat {\bar E}}_\ell $ and the parameters defining ${\hat {\bar H}}_{\nu}$. We begin with Eq.~(\ref{Wdef3}), which expresses $ F^{ab}_{\ell}$ in terms of an operator ${\hat {\bar W}}[\ell]$ and most easily accomplish our goal by making  the following rearrangement of terms in ${\hat {\bar W}}[\ell]$,
\beq
\label{pdef4}
{\hat {\bar W}}[\ell] &=&  U {\hat {\bar H}}^2_{0v} U^{\dag} - U {\hat {\bar H}}_{0v}U^{\dag}~ \Sigma[\ell] \nonumber \\
&+&  U {\hat {\bar H}}_{0v} U^{\dag} {\hat V} + {\hat V} U  {\hat {\bar H}}_{0v} U^{\dag} \nonumber \\
&+& {\hat V}^2  -  {\hat V}~  \Sigma[\ell] + {\bf 1}~ \Pi[\ell] ~.
\eeq
Here,  $\Sigma[\ell]$ is the sum, and $\Pi[\ell]$ the product, of two EV as defined in Eq.~(\ref{SPD},\ref{DelEdef}).

It is useful to recall that  $\Sigma[\ell]$ and $\Pi[\ell]$, defined in this fashion,  depend on $\ell$ entirely through ${\hat {\bar E}}_\ell$. 
In particular~\cite{OS}, from  Eq.~(\ref{Sig1}),  
\beq
\label{Sig2x}
\Sigma[\ell] &=& -a - {\hat {\bar E}}_\ell ~,
\eeq
and, from  Eq.~(\ref{Pi1}),  
\beq
\label{Pi2x}
\Pi[\ell] &=& b + a {\hat {\bar E_\ell}} + {\hat {\bar E_\ell}}^2    ~,
\eeq
with $a,b$ given in Eq.~(\ref{abcvals}).

Explicit expressions for ${\hat {\bar W}}[\ell]$ are easily found in terms of $H_\nu$ using  Eq.~(\ref{pdef4}), from which it follows that  
its entire dependence on $\delta_{cp}$ occurs through three operators, $W_0[\ell]$,  $W_{cos}[\ell]$ and $W_{sin}[\ell]$,
\beq
\label{Pstru}
{\hat {\bar W}}[\ell] &=& {\hat {\bar W}}_0[\ell] + \cos{\delta_{cp}} {\hat {\bar W}}_{cos}[\ell] \nonumber \\
&+& i \sin{\delta_{cp}} {\hat {\bar W}}_{sin}[\ell] ~.
\eeq
with ${\hat {\bar W}}_0[\ell]$, ${\hat {\bar W}}_{cos}[\ell]$ and ${\hat {\bar W}}_{sin}[\ell]$ real and independent of $\delta_{cp}$. Additionally, ${\hat {\bar W}}_{sin}[\ell]$ anti-symmetric, whereas ${\hat {\bar W}}_{cos}[\ell]$ and ${\hat {\bar W}}_0[\ell] $ are symmetric, under exchange of initial and final states.
It follows that $Tr{\hat {\bar W}}_{cos}[\ell]=Tr{\hat {\bar W}}_{sin}[\ell]=0$, and therefore 
\beq
\label{trW}
Tr  {\hat {\bar W}}[\ell]   &=& Tr {\hat {\bar W}}_0[\ell]  ~.
\eeq
Then, from Eq.~(\ref{trW}) and
the definition of $F^{ab}_{\ell}$ in Eq.~(\ref{Wdef}), we find the trace relationship 
\beq
Tr F_\ell &=& \frac{ Tr  {\hat {\bar W}}[\ell]  }{ {\hat {\bar  D}}[\ell] } = 1 ~.
\eeq

\subsubsection{ Analytic Expressions for ${\hat {\bar W}}_{0}[\ell]$ }

For flavor changing transitions, $\nu_i \to \nu_j$, the matrix elements of ${\hat {\bar W}}_0[\ell]$ are found to have the structure  
\beq
<M(\mu)| {\hat {\bar W}}_0[\ell] |M(e)>  &\equiv& {\hat {\bar W}}^{12}_0[\ell] = ~c^{(12)}[\ell] \cos{\theta_{23}} \nonumber \\
<M(\tau)|{\hat {\bar W}}_0[\ell] |M(e)> &\equiv& {\hat {\bar W}}^{13}_0[\ell] = - c^{(12)}[\ell] \sin{\theta_{23}} \nonumber \\
<M(\tau)|{\hat {\bar W}}_0[\ell] |M(\mu)> &\equiv& {\hat {\bar W}}^{23}_0[\ell]  \nonumber \\
&=& ~c^{(23)}[\ell] \sin{2\theta_{23}} ~.
\eeq
We use the convention that a quantity $O^{(ab)}$ written {\it with} parentheses enclosing $ab$ is a number, not a matrix element. As noted, $O^{ab}$ written {\it without} parentheses surrounding $ab$ is a matrix element. This distinction may be obvious in the present context, but later on this distinction may not be so obvious.

Note that ${\hat {\bar W}}^{e\mu}_0[\ell]$ and ${\hat {\bar W}}^{e\tau}_0[\ell] $ are not independent.
The corresponding coefficients $ c^{(ii)}[\ell] $ of for flavor preserving transitions are defined with ${\hat {\bar E}} _{\ell}^2 $ in  Eq.~(\ref{Pi2x}) separated out,
\beq
<M(i)| {\hat {\bar W}}_0[\ell] |M(i)>  &\equiv& {\hat {\bar W}}^{ii}_0[\ell] \nonumber \\
&=&  {\hat {\bar E}}_{\ell}^2 + c^{(ii)}[\ell] ~.
\eeq
The matrix for ${\hat {\bar W}}_0[\ell]$ is thus 
\begin{widetext} 
\beq
\label{W0}
{\hat {\bar W}}_{0}[\ell] &=& {\bf 1}\times {\hat {\bar E}} _{\ell}^2  + \left( \begin{array}{ccc} c^{(11)} [\ell] & c^{(12)} [\ell] \cos\theta_{23}    &- c^{(12)} [\ell] \sin\theta_{23}    \\ c^{(12)} [\ell] \cos\theta_{23}         & c^{(22)} [\ell]    & c^{(23)} [\ell] \sin {2 \theta_{23} } \\ -c^{(12)} [\ell] \sin\theta_{23}    & c^{(23)} [\ell] \sin {2 \theta_{23} } & c^{(33)}[\ell] \end{array} \right ) ~.
\eeq

Writing the dependence of $ c^{(ij)} [\ell] $ on  ${\hat {\bar E}} _{\ell}$ explicitly,
\beq
c^{(ij)} [\ell] &=& c_0^{(ij)}    + c_1^{(ij)} ~{\hat {\bar E}} _{\ell} ~,
\eeq
and expressing the results in terms of the combinations of mixing angles 
\beq
\label{Cpmval2}
C_1^{(\pm)} &\equiv& \cos^2{ \theta_{12}} \cos^2{ \theta_{23}} \pm \sin^2{ \theta_{23}} \sin^2{ \theta_{12}}  \sin^2{ \theta_{13}} \nonumber \\
C_2^{(\pm)} &\equiv& \cos^2{ \theta_{12}} \pm \sin^2{ \theta_{12}}  \sin^2{ \theta_{13}} ~,
\eeq
where $C_2^{(\pm)}$ was defined earlier in  defined in Eq.~(\ref{Cbval}), we find the following exact results.  For the flavor-changing transitions, 
\beq
\label{a1a3vals0}
c^{(12)}_0 &=& - c^{(12)}_1 = -\frac{\alpha}{2} \cos{\theta_{13} } \sin { 2 \theta_{12}  } \nonumber \\
c^{(23)}_0 &=& - \frac{\alpha}{2} ( \sin^2{\theta_{12} } - \cos^2{\theta_{12} } \sin^2{\theta_{13}} ) -  \frac{{\hat A} }{2}  (\cos^2{ \theta_{13}} - \alpha C_2^{(-)} )
\nonumber \\
c^{(23)}_1 &=&  \frac{1}{2} ( \cos^2{\theta_{13} } -  \alpha C_2^{(-)} ) ~;
\eeq
and, for  the flavor-preserving transitions,
\beq
c_0^{(11)} &=&  \alpha   \cos^2{\theta_{12}} \cos^2{\theta_{13}} \nonumber \\
c_1^{(11)} &=& 
-\cos^2{\theta_{13}} -   \alpha C_2^{(+)}  ~,
\eeq
\beq
c^{(22)}_0 &=&   \alpha (1 - \sin^2{\theta_{23}} \cos^2{\theta_{13}} - C_2^{(+)})  +{\hat A} ( \cos^2{\theta_{13}} \cos^2{\theta_{23}} + \alpha 
(C_2^{(+)}- C_1^{(+)})) \nonumber \\
c^{(22)}_1 &=& -1 + \cos^2{\theta_{13}}\sin^2{\theta_{23}}    - \alpha (1- C_1^{(+)}) -  {\hat A}    ~.
\eeq
The coefficients $ c^{(33)} $ follow  from $ c^{(22)}$ by the exchange $\sin{\theta_{23} }  \leftrightarrow \cos{\theta_{23} }$. 
Consequently, there are only four independent matrix elements for ${\hat {\bar W}}_0[\ell]$, two diagonal and two off-diagonal. With the eigenvalues independent of $\theta_{23}$, it follows that 
\beq
\label{W0sym}
{\hat {\bar W}}^{\tau\tau}_0[\ell] &=& {\hat {\bar W}}^{\mu\mu}_0[\ell] |_{ \sin{\theta_{23} }  \leftrightarrow \cos{\theta_{23} } } ~.
\eeq

\subsubsection{ Analytic Expressions for ${\hat {\bar W}}_{sin}[\ell]$ and ${\hat {\bar W}}_{cos}[\ell]$ }

The matrices for ${\hat {\bar W}}_{sin}[\ell]$ and ${\hat {\bar W}}_{cos}[\ell]$ have the following simple structure 
\beq
\label{Wsin}
{\hat {\bar W}}_{sin}]\ell] &=& \left( \begin{array}{ccc} 0 & -a^{(12)}[\ell] \sin\theta_{23}    & -a^{(12)}[\ell] \cos\theta_{23}  \\ a^{(12)}[\ell] \sin\theta_{23}     & 0  &  a^{(23)}[\ell] \\ a^{(12)}[\ell]\cos\theta_{23}      & - a^{(23)}[\ell]   & 0   \end{array} \right ) 
\eeq
and
\beq
\label{Wcos}
{\hat {\bar W}}_{cos} [\ell] &=& \left( \begin{array}{ccc} 0 & a^{(12)}[\ell]\sin\theta_{23}    & a^{(12)}[\ell]\cos\theta_{23}  \\ a^{(12)}[\ell]\sin\theta_{23}     & - a^{(23)}[\ell]\sin 2\theta_{23}  & - a^{(23)}[\ell] \cos 2\theta_{23} \\  a^{(12)}[\ell] \cos\theta_{23}      & - a^{(23)}[\ell] \cos 2\theta_{23}   & a^{(23)}[\ell]  \sin 2\theta_{23}   \end{array} \right ) ~,
\eeq
\end{widetext}
where
\beq
\label{aelldef}
a^{(23)}[\ell] &=& a^{(23)}_0 + a^{(23)}_1 ~ {\hat {\bar E}} _{\ell} \nonumber \\
a^{(12)}[\ell] &=& a^{(12)}_0 + a^{(12)}_1 ~ {\hat {\bar E}} _{\ell} ~.
\eeq
We find the exact results
\beq
\label{a1a3vals}
a^{(23)}_0 &=& -\frac{ {\hat A}   \alpha }{2} \sin{2 \theta_{12} } \sin{ \theta_{13} } -\frac{ \alpha}{2} \sin{2 \theta_{12} } \sin{ \theta_{13} } \nonumber \\
a^{(23)}_1 &=& \frac{\alpha }{2} \sin{2 \theta_{12} } \sin{\theta_{13} }  \nonumber \\
a^{(12)}_0 &=& - \frac{\alpha }{2} \cos^2{\theta_{12} }  \sin{2 \theta_{13} } \nonumber \\
a^{(12)}_1 &=& \frac{1}{2} \sin{2 \theta_{13} } - \frac{\alpha}{2} \sin^2 {\theta_{12} }  \sin{2 \theta_{13} } ~.
\eeq

\subsubsection{ Interconnections among  ${\hat {\bar W}}_{0}[\ell]$, ${\hat {\bar W}}_{sin}[\ell]$, and ${\hat {\bar W}}_{cos}[\ell]$ }

Significant simplifications arise from interconnections among $W_0[\ell]$,  $W_{cos}[\ell]$, and $W_{sin}[\ell]$.
One of these follows from Eqs.(\ref{Wsin},\ref{Wcos}), which show that the matrix elements of ${\hat {\bar W}}_{cos}[\ell]$ and ${\hat {\bar W}}_{sin}[\ell]$ are proportional, namely
\beq
\label{a:screl}
&&<M(a)| {\hat {\bar W}}_{sin}[\ell] |M(b)> \nonumber \\
&=&  f_{ab} <M(a)| {\hat {\bar W}}_{cos}[\ell] |M(b)> ~,
\eeq
with the constants of proportionality $f_{ab}$ independent of $\ell$,
\beq
\label{a:screlc}
f_{ab} &=& \left( \begin{array}{ccc} 0 &  -1  & -1  \\  1 & 0 & -\cos^{-1}{2 \theta_{23}} \\ 1  &  \cos^{-1}{2 \theta_{23}} & 0  \end{array} \right ) ~.
\eeq
Another follows from Eqs.~(\ref{a1a3vals0},\ref{a1a3vals}), from which it follows that 
\beq
\label{ABval}
a^{(12)}_1 c^{(12)}_0 - a^{(12)}_0 c^{(12)}_1 &=& 2 K \nonumber \\
a^{(23)}_1 c^{(23)}_0 - a^{(23)}_0 c^{(23)}_1 &=& - K ~.
\eeq
The quantity $K$ is
\beq
\label{Kdef}
K &=& -\frac{\alpha(1-\alpha) }{8} \nonumber \\
&\times& \cos{\theta_{13}} \sin{2\theta_{12}} \sin{2\theta_{13}} ~.
\eeq

\subsection{ Expressions for $ S(T)$ in terms of $\delta_{cp}$ }

The dependence of $ S(T)$ on the CP violating phase $\delta_{cp}$ is very simple and follows from  Eqs.~(\ref{lagrf5},\ref{Pstru}), noting that $ F^{ab}_{\ell} = {\hat {\bar W}}^{ab}[\ell] / {\hat {\bar D}}[\ell] $, Eq.~(\ref{Fabal1}).  We thus find,  
\beq
\label{lagrf6}
Re[ S^{ab}(t',t) ] &=&  \delta_{ab} -  2 \sum_{\ell}  
\frac{ {\hat {\bar W}}^{ab}_0[\ell]} { {\hat {\bar D}}[\ell] }\sin^2{ {\hat{\bar E}}_{\ell} \Delta_L  } \nonumber \\
&-&  2 \cos{\delta_{cp}} \sum_{\ell}  
\frac{ {\hat {\bar W}}^{ab}_{cos}[\ell] } { {\hat {\bar D}}[\ell] }\sin^2{ {\hat{\bar E}}_{\ell} \Delta_L  } \nonumber \\
&+& \sin\delta_{cp} \sum_{\ell}  \frac{ {\hat {\bar W}}^{ab}_{sin}[\ell]  } { {\hat {\bar D}}[\ell] }\sin{ 2{\hat{\bar E}}_{\ell} \Delta_L  }    ~,
\eeq
and
\beq
Im[ S^{ab}(t',t) ] &=&  -  2 \sin\delta_{cp} \sum_{\ell}  
\frac{  {\hat {\bar W}}^{ab}_{sin}[\ell] } { {\hat {\bar D}}[\ell] }  \sin^2{ {\hat{\bar E}}_{\ell} \Delta_L  } \nonumber \\
&-& \sum_{\ell} \frac{  {\hat {\bar W}}^{ab}_0[\ell] } { {\hat {\bar D}}[\ell] }  \sin{ 2{\hat{\bar E}}_{\ell} \Delta_L  } \nonumber \\
&-& \cos{\delta_{cp}} \sum_{\ell} \frac{  {\hat {\bar W}}^{ab}_{cos}[\ell] } { {\hat {\bar D}}[\ell] }  \sin{ 2{\hat{\bar E}}_{\ell} \Delta_L  } ~.
\eeq
The dependence of $S(L)$ on ${\hat {\bar E}} _{\ell}$ and the remaining parameters of $H_\nu$ is given analytically through Eqs.~(\ref{W0},\ref{Wsin},\ref{Wcos}). 

\section {Oscillation Probability  $\mathcal{P}(\nu_a  \rightarrow \nu_b)$ in our Hamiltonian formulation }
\label{PartOscProb} 

The neutrino oscillation probabilities are obtained directly from our expression for the time-evolution operator, Eq.~(\ref{lagrf}).
In the high-energy limit, the oscillation probability in Eq.~(\ref{oscprob}), expressed in terms of $Re[F^{ab}_{\ell\ell'}]$ and $Im[F^{ab}_{\ell\ell'}]$, is
\beq
\label{a:Pmab2}
&&\mathcal{P}(\nu_a  \rightarrow \nu_b) =  2\sum_{\ell > \ell'} Im[F^{ab}_{\ell\ell'}] \sin{ (2 \Delta {\hat{\bar E}}_{\ell\ell'}
\Delta_L )} \nonumber \\
&+&  2\sum_{\ell>\ell'} Re [F^{ab}_{\ell \ell'}] \cos{( 2 \Delta {\hat{\bar E}}_{\ell\ell'}  \Delta_L ) }  ~,
\eeq
where $1 \leq \ell \leq 3$, where $\Delta_L$ was defined in Eq.~(\ref{Deldeff}), and $\Delta {\hat{\bar E}}_{\ell\ell'} $ in Eq.~(\ref{DelEdef}).

Note that $Re [F^{ab}_{\ell \ell'}]$ satisfies a normalization condition 
\beq
\label{normF1}
\sum_{\ell\ell'} Re F^{ab}_{\ell\ell'}  &=& \delta_{ab} ~,
\eeq
found by evaluating Eq.~(\ref{oscprob})  at $t = t'$ and recognizing that $\mathcal{P}(\nu_a  \rightarrow \nu_b)$ is a real number.
Taking Eq.~(\ref{normF1}) into account and using the identity $1  - \cos{ 2 \beta }  \equiv 2\sin^2{ \beta}$, we find an equivalent  expression 
\beq
\label{a:poprob}
&&\mathcal{P}(\nu_a  \rightarrow \nu_b) = \delta_{ab} + 2\sum_{\ell >\ell'} Im[ F^{ab}_{\ell\ell'}] \nonumber \\
&\times& \sin{ ( 2  \Delta {\hat{\bar E}}_{\ell\ell'}
\Delta_L )} \nonumber \\
&-& 4\sum_{\ell >\ell'} Re F^{ab}_{\ell \ell'} \sin^2{ (\Delta {\hat{\bar E}}_{\ell\ell'} \Delta_L ) }  ~,
\eeq
that  bears  a striking similarity to the familiar vacuum expression with  $F^{ab}_{\ell\ell'}$ playing a role analogous to  $J^{ab}_{\ell\ell'}$ (as, for example, in Eq.~(1) of Ref.~\cite{f}). Equations~(\ref{a:Pmab2},\ref{a:poprob}) make use of  the fact that a probability is purely real.

\subsection {Properties of $F^{ab}_{\ell\ell'}$ }

It follows from Eq.~(\ref{Pmab1}) and the observation that the energy of a neutrino or antineutrino in matter is independent of $\delta_{cp}$ that $F^{ab}_{\ell\ell'}$ is symmetric under the simultaneous exchange of $a,b$ and $\ell,\ell'$,
\beq
\label{Jsym0}
F^{ab}_{\ell\ell'} &=& F^{ba}_{\ell' \ell} ~. 
\eeq
Using in addition the Hermiticity of ${\hat {\bar W}}[\ell]$, Eq.~(\ref{hermitp}) we find the reflection symmetry 
\beq
F^{ba}_{\ell\ell'} &=& F^{ab*}_{\ell\ell'} \nonumber \\
F^{ab}_{\ell'\ell} &=& F^{ab*}_{\ell\ell} ~.
\eeq
From this, along with Eq.~(\ref{Jsym0}), it follows  that $Im F^{ab}_{\ell\ell'}$ is antisymmetric under the exchange of either  $(a,b)$ or $(\ell,\ell')$, 
\beq 
\label{Jsym1}
Im F^{ab}_{\ell \ell'} &=& -Im F^{ab}_{\ell' \ell} = -Im F^{ba}_{\ell\ell' } ~,
\eeq
whereas $ReF^{ab}_{\ell\ell'}$ is symmetric,
\beq
\label{Jsym2}
Re F^{ab}_{\ell \ell'} &=& Re F^{ab}_{\ell'\ell} =  Re F^{ba}_{\ell\ell' } ~.
\eeq

There are no symmetries connecting the $F^{ab}_{\ell\ell'}$ of neutrinos to those of antineutrinos because these energies are, in general, different.  This is not the case, however, in the vacuum for theories invariant under CPT.

\subsection{ General Expressions for $ F^{ab}_{\ell\ell'}$ }

The quantity $ F^{ab}_{\ell  \ell'} $ is most easily obtained from  $ w^{ab}_{\ell  \ell'}  $,
\beq
\label{Pmab1w}
w^{ab}_{\ell \ell'} &\equiv&   <M(b)| {\hat {\bar W}}[\ell] |M(a)> \nonumber \\
&\times& <M(b)|{\hat {\bar W}}[\ell'] |M(a)>^*  ~.
\eeq
Analytic formulae for $ w^{ab}_{\ell,\ell'} $ are easily obtained in terms of the parameters of $H_\nu$ 
using ${\hat {\bar W}} [\ell] $ given  in Eq.~(\ref{Pstru}). Equation~(\ref{Pmab1}) then gives $ F^{ab}_{\ell  \ell'} $ as
\beq
\label{Pmab2}
F^{ab}_{\ell  \ell'} &=& \frac{ w^{ab}_{\ell  \ell'}  } {{\hat {\bar D}}[\ell]   {\hat {\bar D}}[\ell']  } ~.
\eeq

Because ${\hat {\bar W}} [\ell] $ consists of three terms, one proportional to $\sin\delta_{cp}$, one proportional to $\cos\delta_{cp}$, and one independent of $\delta_{cp}$, the dependence of $ w^{ab}[\ell,\ell'] $  on $\delta_{cp}$  can be expressed {\it a priori}  through the five operators,  
\beq
\label{a:Pmabs2}
w^{ab}_{\ell\ell'} &=& w_{0\ell\ell'} ^{ab} + \cos{\delta_{cp}} w_{cos\ell\ell'}^{ab} \nonumber \\
&+& \cos^2{\delta_{cp}} w_{cos^2\ell\ell'}^{ab} 
+ i( \sin{\delta_{cp}} w_{sin\ell\ell'}^{ab} \nonumber \\
&+& \sin{\delta_{cp}}\cos{\delta_{cp}}w^{ab}_{sin\times cos\ell\ell'} ~, 
\eeq
each of which is uniquely determined by Eqs.~(\ref{Pstru},\ref{Pmab1w}). Clearly, just as for the ${\hat {\bar W}}_i[\ell]$, the matrix elements of $w_i[\ell,\ell']$ are real and independent of $\delta_{cp}$.

Using the hermiticity of ${\hat {\bar W}}_{0}[\ell] $ and ${\hat {\bar W}}_{cos}[\ell] $, and the anti-hermiticity of ${\hat {\bar W}}_{sin}[\ell] $, we then find
\beq
\label{a:PFabalbe}
w^{ab}_{0\ell\ell'} &\equiv& {\hat {\bar W}}^{ab}_0[\ell] {\hat {\bar W}}^{ab}_0[\ell'] + {\hat {\bar W}}^{ab}_{sin}[\ell] {\hat {\bar W}}^{ab}_{sin}[\ell'] \nonumber \\
w_{sin\ell\ell'}^{ab} &\equiv& {\hat {\bar W}}^{ab}_{sin}[\ell] {\hat {\bar W}}^{ab}_0[\ell'] - {\hat {\bar W}}^{ab}_0[\ell] {\hat {\bar W}}^{ab}_{sin}[\ell']
\nonumber \\
w^{ab}_{cos\ell \ell'} &\equiv& {\hat {\bar W}}^{ab}_{cos}[\ell] {\hat {\bar W}}^{ab}_0[\ell'] + {\hat {\bar W}}^{ab}_0[\ell] {\hat {\bar W}}^{ab}_{cos}[\ell'] \nonumber \\
w_{cos^2\ell\ell'}^{ab} &\equiv& {\hat {\bar W}}^{ab}_{cos}[\ell] {\hat {\bar W}}^{ab}_{cos}[\ell'] - {\hat {\bar W}}^{ab}_{sin}[\ell] {\hat {\bar W}}^{ab}_{sin}[\ell'] \nonumber \\
w^{ab}_{sin\times cos\ell\ell'} &\equiv& {\hat {\bar W}}^{ab}_{sin}[\ell] {\hat {\bar W}}^{ab}_{cos}[\ell'] \nonumber \\
&~&~~~~~~~~~ - {\hat {\bar W}}^{ab}_{cos}[\ell] {\hat {\bar W}}^{ab}_{sin}[\ell']~.
\eeq 
Note that Eqs.~(\ref{a:screl},\ref{a:screlc}) require that $w_{sin\times cos}$ vanish identically, 
\beq
\label{a:Psincos}
w^{ab}_{sin\times cos} &=& 0 ~,
\eeq
so this term need not be considered further.  Note also that the dependence of  $ w^{ab}_{i\ell\ell'} $ on $\ell$ and $\ell'$ arises entirely from the eigenvalues ${\hat {\bar E}} _{\ell}$ and ${\hat {\bar E}} _{\ell'}$, as ${\hat {\bar W}}^{ab}[\ell]$ depends on $\ell$ entirely through ${\hat {\bar E}} _{\ell}$.
Finally, we will find it useful to define $ w_{i}^{ab} [\ell]  $ in analogy to ${\hat {\bar W}}[\ell] $ 
in Eq.~(\ref{Wdef3}),
\beq
w_{i}^{ab} [1]  &\equiv& w_{i}^{ab}  [3,2] \nonumber \\
w_{i}^{ab}  [2]  &\equiv& w_{i}^{ab}  [3,1] \nonumber \\
w_{i}^{ab}  [3]  &\equiv& w_{i}^{ab}  [2,1] ~.
\eeq
These are the only three $ w_{i}^{ab}  [\ell,\ell']$ needed because of the restriction $\ell >\ell'$ in Eq.~(\ref{a:poprob}).

\subsection{ Analytic expressions for $ w_{sin}^{ab} [\ell,\ell'] $ }

From the symmetries of ${\hat {\bar W}} ^{ab}_{0}[\ell]$, ${\hat {\bar W}}^{ab}_{cos}[\ell]$, and ${\hat {\bar W}}^{ab}_{sin}[\ell]$, we see from Eq.~(\ref{a:PFabalbe}) that  $w^{ab}_{sin\ell\ell'}$ is 
odd under the exchange of either $a,b$ or $\ell,\ell'$. The term $w^{ab}_{sin\ell,\ell'}$ therefore 
vanishes for $a=b$ [and for  $\ell = \ell'$], but again recall that restrictions on the sums is such that $w^{ab}_{\ell\ell'}$ contributes only for $\ell > \ell'$. 

Using the general results in Eqs.~(\ref{W0},\ref{Wsin},\ref{Wcos}), we find $ w_{sin\ell\ell'}^{ab} $ given in Eq.~(\ref{a:PFabalbe}) is, as a matrix,
\beq
\label{wsineval}
w_{sin\ell\ell'} &=& \frac{1}{2} \sin{2\theta_{23}} \left( \begin{array}{ccc} 0 &  A &  -A \\ -A  & 0  &  B    \\  A     &  -B    & 0   \end{array} \right ) ~,
\eeq
where
\beq
A  &=& -( a^{(12)}[\ell] c^{(12)}[\ell'] - c^{(12)}[\ell] a^{(12)}[\ell']) \nonumber \\
&=& -(a^{(12)}_1 c^{(12)}_0 - a^{(12)}_0 c^{(12)}_1) \Delta {\hat{\bar E}}_{\ell\ell'}
\eeq
and
\beq
B   &=& 2 (a^{(23)}[\ell] c^{(23)}[\ell'] - c^{(23)}[\ell]  a^{(23)}[\ell']) \nonumber \\
&=&   2(a^{(23)}_1 c^{(23)}_0 - a^{(23)}_0 c^{(23)}_1) \Delta {\hat{\bar E}}_{\ell\ell'}
 ~.
\eeq
We see from Eq.~(\ref{ABval}) that that $A=B$, so from Eq.~(\ref{wsineval}) we find the following simple expression, 
\beq
\label{wsineva2}
w^{ab} _{sin\ell\ell'} &=&  K  \sin{2\theta_{23}} \Delta {\hat{\bar E}}_{\ell\ell'} \epsilon^{ab}_{\sin} ~,
\eeq
and correspondingly
\beq
\label{wsineva2c}
w^{ab} _{sin}[\ell] &=&  K  \sin{2\theta_{23}} \Delta {\hat{\bar E}}[\ell] \epsilon^{ab}_{\sin} ~,
\eeq

where  $\epsilon_{\sin}$ is the anti-symmetric matrix 
\beq
\label{epsidef}
\epsilon_{sin} &\equiv& \left( \begin{array}{ccc} 0 &  1 &  -1  \\  -1   & 0 & 1 \\     1 &  -1    & 0 \end{array} \right) ~.
\eeq

\subsection{ Analytic expressions for $ w^{ab}_{cos^2} $, $ w^{ab}_{cos} $, and  $ w^{ab}_0 $ }
\label{analw}
 
We also see from the symmetries of ${\hat {\bar W}} ^{ab}_{0}[\ell]$, ${\hat {\bar W}}^{ab}_{cos}[\ell]$, and ${\hat {\bar W}}^{ab}_{sin}[\ell]$, that $w^{ab}_{0\ell\ell'}$, $w^{ab}_{cos\ell\ell'}$, and $w^{ab}_{\cos^2\ell\ell'}$ are even under the exchange of $a,b$ and/or $\ell,\ell'$.  
Since the dependence of all $ w^{ab}_{i\ell\ell'} $ on $\ell$ and $\ell'$ arises entirely from the eigenvalues ${\hat {\bar E}} _{\ell}$ and ${\hat {\bar E}} _{\ell'}$, it follows that the terms symmetric in $\ell$ and $\ell'$ must be functions of symmetric combinations of ${\hat {\bar E}} _{\ell}$ and ${\hat {\bar E}} _{\ell'}$.  

There are only three irreducible symmetric functions of ${\hat {\bar E}} _{\ell}$ and ${\hat {\bar E}} _{\ell'}$: a constant, the sum ${\hat {\bar E}} _{\ell}+{\hat {\bar E}} _{\ell'}$, and the product ${\hat {\bar E}} _{\ell}{\hat {\bar E}} _{\ell'}$, just as for ${\hat {\bar W}}[\ell] $ in Eq.~(\ref{pdef4}).
Thus,  we anticipate that the entire dependence of $w^{ab}_0[\ell]$, $w^{ab}_{cos}[\ell]$, and $w^{ab}_{\cos^2}[\ell]$ on $\ell$ will occur through $\Sigma[\ell]$ and $\Pi[\ell]$ in Eqs.~(\ref{Sig2x},\ref{Pi2x}), and hence through ${\hat {\bar E}} _{\ell}$, as for ${\hat {\bar W}}[\ell] $.

In the following discussion, it is important to recall our convention that a quantity, such as  $w^{(ab)}[\ell]$, written {\it with} parentheses enclosing $ab$ is a number, whereas $w^{ab}[\ell]$ written {\it without} parentheses surrounding $ab$ is the element of a matrix $w[\ell]$,
\beq
w^{ab}[\ell] &\equiv& <M(b)| w[\ell] |M(a)> ~.
\eeq

\subsubsection{ Matrix Elements of  $ w_{cos^2} $ }

Using Eq.~(\ref{a:PFabalbe}) with Eqs.~(\ref{Wsin},\ref{Wcos}), the matrix for
$w_{cos^2} $ is found to be 
\beq
\label{a:Pcos2g}
w_{cos^2}  &=& \sin^2{2\theta_{23}} \nonumber \\
&\times& 
\left( \begin{array}{ccc} 0 & 0 & 0 \\ 0  & w^{(22)}_{cos^2}  & - w^{(22)}_{cos^2}  \\ 0 & - w^{(22)}_{cos^2}  & w^{(22)}_{cos^2}  \end{array} \right ) ~,
\eeq
where
\beq
w^{(22)}_{cos^2\ell\ell'}  &=& a^{(23)}[\ell] a^{(23)}[\ell'] \nonumber \\
&=& a^{(23)2}_0 + a^{(23)}_0 a^{(23)}_1  ({\hat {\bar E}} _{\ell} + {\hat {\bar E}} _{\ell'}) \nonumber \\
&+& a^{(23)2}_1  {\hat {\bar E}} _{\ell}{\hat {\bar E}} _{\ell'} ~.
\eeq 
Correspondingly, 
\beq
w^{(22)}_{cos^2} [\ell]  &=& a^{(23)2}_0 + a^{(23)}_0 a^{(23)}_1~ \Sigma[\ell] \nonumber \\
&+& a^{(23)2}_1~ \Pi[\ell]  ~.
\eeq

Note the somewhat subtle notational distinction between the ${\it operator}$  $ w_{cos^2} $ and the ${\it coefficients }$ $ w^{(ij)}_{cos^2} $ in terms of which it is defined.   The latter is indicated by parentheses that surrounding the superscripts. By contrast, superscripts without parentheses, as in $ w^{ab}_i $,  indicate the transition $\nu_a \rightarrow \nu_b$, ${i.e.}$, $w^{ab}_i \equiv <M(b) | w_i | M(a)>$.

\subsubsection{ Matrix Elements of $ w_{cos} $ }

We find only three independent, non-vanishing matrix elements for  $w_{cos}$, one diagonal element and two off-diagonal elements. 
As a matrix, 
\beq
\label{wcoseva0l} 
&& w_{cos}   = \sin 2\theta_{23} \nonumber \\
&\times& \left( \begin{array}{ccc} 0  & w^{(12)}_{cos}  & - w^{(12)}_{cos}  \\ w^{(12)}_{cos}  & w^{(22)}_{cos}    & w^{(23)}_{cos}   \\ - w^{(12)}_{cos}  & w^{(23)}_{cos}     & w^{(33)}_{cos}         \end{array} \right )~.
\eeq 
The two independent off-diagonal elements,
\beq
w^{(12)}_{cos\ell\ell'}  &=& \frac{1}{2} (a^{(12)}[\ell] c^{(12)} [\ell'] + c^{(12)} [\ell] a^{(12)}[\ell'] ) \nonumber \\
w^{(23)}_{cos\ell\ell'}   &=& -( a^{(23)}[\ell] c^{(23)} [\ell'] 
+ c^{(23)} [\ell] a^{(23)}[\ell'] ) \nonumber \\
&\times& \cos{2\theta_{23}}    ~,
\eeq
follow immediately from the structure of ${\hat {\bar W}}^{ab}_0[\ell'] $ and  ${\hat {\bar W}}^{ab}_{cos}[\ell]$ given in Eqs.~(\ref{W0},\ref{Wcos}), respectively. 

Taking $ w^{(22)}_{cos} $ from Eqs.~(\ref{W0},\ref{Wcos},\ref{a:PFabalbe}), we obtain
\beq
w^{(22)}_{cos\ell\ell'} &=& - a^{(23)}[\ell] ({\hat {\bar E}} _{\ell'}^2    + c^{(22)} [\ell'] ) \nonumber \\
&-& ({\hat {\bar E}} _{\ell}^2    + c^{(22)} [\ell] ) a^{(23)}[\ell'] ~.
\eeq
Similarly, for the diagonal matrix element $D$, 
\beq
w^{(33)}_{cos\ell\ell'}  
&=&  a^{(23)}[\ell] ({\hat {\bar E}} _{\ell'}^2    + c^{(33)} [\ell'] ) \nonumber \\
&+& ({\hat {\bar E}} _{\ell}^2    + c^{(33)} [\ell] ) a^{(23)}[\ell'] ~.
\eeq 
The element $ w^{(11)}_{cos\ell\ell'}$ vanishes as a consequence of ${\hat {\bar W}}^{(11)}_{cos}[\ell]=0$.

Correspondingly, for the off-diagonal matrix elements we find,
\beq
w^{(12)}_{cos} [\ell] &=& a^{(12)}_0 c^{(12)}_0 + \frac{1}{2} ( a^{(12)}_1 c^{(12)}_0 + a^{(12)}_0 c^{(12)}_1 ) \nonumber \\
&\times& \Sigma[\ell] + a^{(12)}_1 c^{(12)}_1 \Pi[\ell] \nonumber \\
w^{(23)}_{cos} [\ell] &=& -2 \cos{2\theta_{23}}  a^{(23)}_0 c^{(23)}_0  \nonumber \\
&-& \cos{2\theta_{23}} ( a^{(23)}_1 c^{(23)}_0 + a^{(23)}_0 c^{(23)}_1 ) \nonumber \\ 
&\times& \Sigma[\ell] 
- 2 \cos{2\theta_{23}} a^{(23)}_1 c^{(23)}_1 ~\Pi[\ell] ~,
\eeq
which simplify somewhat by using Eqs.~(\ref{ABval},\ref{Kdef}).
For the diagonal matrix elements,
\beq
w^{(22)}_{cos} [\ell] &=& -2 a^{(23)}_0 c^{(22)}_0 - ( a^{(23)}_1 c^{(22)}_0 + a^{(23)}_0 c^{(22)}_1) \nonumber \\
&\times& \Sigma[\ell] +  2( a^{(23)}_0  - a^{(23)}_1 c^{(22)}_1)~ \Pi[\ell] \nonumber \\
&-& a^{(23)}_1 \Sigma[\ell] ~\Pi[\ell] - a^{(23)}_0 \Sigma[\ell]^2 
\eeq
and
\beq
w^{(33)}_{cos} [\ell]  &=& 2 a^{(23)}_0 c^{(33)}_0 + ( a^{(23)}_1 c^{(33)}_0 + a^{(23)}_0 c^{(33)}_1) \nonumber \\
&\times& \Sigma[\ell] -  2( a^{(23)}_0  - a^{(23)}_1 c^{(33)}_1)~ \Pi[\ell] \nonumber \\
&+& a^{(23)}_1 \Sigma[\ell] ~\Pi[\ell] + a^{(23)}_0 \Sigma[\ell]^2 
~.
\eeq 
The matrix elements  $ w^{(22)}_{cos}[\ell]    $ and $ w^{(33)}_{cos}[\ell]    $ are not independent.  The reason is that 
$ w_{cos}^{(33)} [\ell]$ may be found from $ w_{cos}^{(22)} [\ell]$ by making the replacement $\sin{\theta_{23}} \leftrightarrow \cos{\theta_{23}} $ [Eq.~(\ref{W0sym})] and by flipping the overall sign.  Recall that under this replacement, $ c^{(22)} \leftrightarrow c^{(33)} $ and that $ a^{(23)} $ is independent of $\sin{\theta_{23} }$.

\subsubsection{ Matrix Elements of $ w_{0} $ }

We find four independent, non-vanishing matrix elements for  $w_{0}$, two diagonal and two off-diagonal elements. 
As a matrix, 
\beq
\label{w0eva0l}
&& w_{0}   =  \left( \begin{array}{ccc} w_{0}^{(11)}     & w_{0}^{(12)} & w_{0}^{(13)} \\ w_{0}^{(12)} & w_{0}^{(22)}   & w_{0}^{(23)} \\ w_{0}^{(13)} & w_{0}^{(23)}   & w_{0}^{(33)} \end{array} \right )~.
\eeq
The off-diagonal elements,
\beq
w_{0\ell\ell'}^{(12)} &=& c^{(12)} [\ell] c^{(12)} [\ell'] \cos^2\theta_{23}   \nonumber \\
&+&  a^{(12)}[\ell] a^{(12)}[\ell'] \sin^2{\theta_{23} } \nonumber \\
w_{0\ell\ell'}^{(13)} &=& c^{(12)} [\ell] c^{(12)} [\ell'] \sin^2\theta_{23}   \nonumber \\
&+&  a^{(12)}[\ell] a^{(12)}[\ell'] \cos^2{\theta_{23} } \nonumber \\
w_{0\ell\ell'}^{(23)} &=&  c^{(23)}[\ell] c^{(23)}[\ell'] \sin^2 2 \theta_{23}  \nonumber \\
&+&  a^{(23)}[\ell] a^{(23)}[\ell'] ~,
\eeq
follow immediately from the structure of ${\hat {\bar W}}^{ab}_0[\ell'] $ and  ${\hat {\bar W}}^{ab}_{sin}[\ell]$ given in Eqs.~(\ref{W0},\ref{Wsin}), respectively.
The diagonal elements $ w^{(nn)}_{0} $ are
found from Eqs.~(\ref{W0},\ref{a:PFabalbe}),
\beq
w_{0\ell\ell'}^{(nn)} &=& ( {\hat {\bar E}} _{\ell}^2    + c^{(nn)} [\ell] ) \nonumber \\
&\times& ( {\hat {\bar E}} _{\ell'}^2    + c^{(nn)} [\ell'] ) ~.
\eeq

Correspondingly, for the off-diagonal matrix elements,
\beq
w_{0}^{(12)} [\ell] &=& c^{(12)2}_0 \cos^2{\theta_{23}} + a^{(12)2}_0 \sin^2{\theta_{23}} \nonumber \\
&+& (c^{(12)}_0 c^{(12)}_1 \cos^2{\theta_{23}} + a^{(12)}_0 a^{(12)}_1 \sin^2{\theta_{23}} ) \Sigma[\ell] \nonumber \\ 
&+& ( c^{(12)2}_1 \cos^2{\theta_{23}}  +  a^{(12)2}_1 \sin^2{\theta_{23}})~ \Pi[\ell] \nonumber \\
w_{0}^{(13)} [\ell] &=& c^{(12)2}_0 \sin^2{\theta_{23}} + a^{(12)2}_0 \cos^2{\theta_{23}} \nonumber \\
&+& (c^{(12)}_0 c^{(12)}_1 \sin^2{\theta_{23}} + a^{(12)}_0 a^{(12)}_1 \cos^2{\theta_{23}} ) \Sigma[\ell] \nonumber \\ 
&+& ( c^{(12)2}_1 \sin^2{\theta_{23}}  +  a^{(12)2}_1 \cos^2{\theta_{23}})~ \Pi[\ell] \nonumber \\
w_{0}^{(23)} [\ell] &=& c^{(23)2}_0 \sin ^2{2 \theta_{23}} + a^{(23)2}_0  \nonumber \\
&+& ( a^{(23)}_0 a^{(23)}_1 + c^{(23)}_0 c^{(23)}_1 \sin ^2{2 \theta_{23}}  ) ~\Sigma[\ell] \nonumber \\ 
&+& ( c^{(23)2}_1 \sin^2{\theta_{23}}  + a^{(23)2}_1  )~ \Pi[\ell] ~.
\eeq
For the diagonal matrix elements,
\beq
w_{0}^{(nn)} [\ell] &=& c^{(nn)2}_0 - (2   c^{(nn)}_0 - c^{(nn)2}_1 )~\Pi[\ell] \nonumber \\
&+&  c^{(nn)}_0 c^{(nn)}_1 ~ \Sigma[\ell] + c^{(nn)}_1 \Sigma[\ell] ~ \Pi[\ell] \nonumber \\
&+& \Pi[\ell]^2  ~.
\eeq 
As for $ w_{cos}$, the matrix elements  $ w^{(22)}_{0}[\ell]$ and $ w^{(33)}_{0}[\ell]$ are not independent.  The same is true for $ w_{0}^{(13)} [\ell]$ and $ w_{0}^{(12)} [\ell]$.  In both cases the matrix elements may be found from one another by making the replacement $\sin{\theta_{23}} \leftrightarrow \cos{\theta_{23}} $ [Eq.~(\ref{W0sym})].   Recall that under such a replacement $ c^{(22)} \leftrightarrow c^{(33)} $ and that $ c^{(12)} $  and $ a^{(23)} $ are independent of $\sin{\theta_{23} }$.

Writing the dependence of the coefficients $ w_i^{(mn)} [\ell] $ on  ${\hat {\bar E}} _{\ell}$ explicitly,
\beq
\label{wine}
w_i^{(mn)} [\ell] &=& w_{i;0}^{(mn)}    + w_{i;1}^{(mn)} ~{\hat {\bar E}} _{\ell} \nonumber \\
&+& w_{i;2}^{(mn)} ~{\hat {\bar E}} _{\ell}^2 ~.
\eeq

\subsection{ Analytic expression for $ Im[F^{ab}_{\ell\ell'}] $ and $ Re[ F^{ab}_{\ell\ell'}] $ }

A general form for the real and imaginary parts of  $F^{ab}_{\ell\ell'}$ in terms of the four operators, ($w_0$, $w_{sin}$, $w_{cos}$, $w_{cos^2}$) is found from  Eq.~(\ref{Pmab1}) and Eq.~(\ref{a:Pmabs2}). For $Im[F^{ab}_{\ell\ell'}]$,
using Eq.~(\ref{wsineva2}), we find
\beq
\label{Pmabs2aI}
Im[F^{ab}_{\ell \ell'}] &=& \sin{\delta_{cp}} \frac{ w_{sin\ell\ell'}^{ab} }{ {\hat {\bar D}}[\ell]   {\hat {\bar D}}[\ell']    } \nonumber \\
&=&   \frac{1}{2} \sin{\delta_{cp} }  \sin{2\theta_{23}} K \epsilon^{ab}_{\sin} \nonumber \\
&\times& \frac{ \Delta {\hat{\bar E}}_{\ell\ell'}  }{ {\hat {\bar D}}[\ell]   {\hat {\bar D}}[\ell']  } ~.
\eeq
Then, using the easily-verified result
\beq
\label{a:rate}
&&\frac{  \Delta {\hat{\bar E}}_{\ell\ell'} }{ {\hat {\bar D}}[\ell]   {\hat {\bar D}}[\ell']  } =  \frac{ \epsilon^{\ell\ell'}_{\sin} } {  {\hat {\bar D}} } ~,
\eeq
we find
\beq
\label{ImFgen2}
Im F^{ab}_{\ell\ell'} &=& -\sin{\delta_{cp}} \epsilon_{sin}^{ab} \epsilon_{sin}^{\ell\ell' } \frac{\alpha(1-\alpha) }{ 8{\hat {\bar D}} }  \cos{\theta_{13}} \nonumber \\
&\times& \sin{2\theta_{12}} \sin{2\theta_{13}} \sin{2\theta_{23}}  ~.
\eeq
For $Re[F^{ab}_{\ell\ell'}]$ we find   
\beq
\label{Pmabs2aR}
Re[F^{ab}_{\ell \ell'}] &=& \frac{ w_{0\ell\ell'}^{ab} }{ {\hat {\bar D}}[\ell]   {\hat {\bar D}}[\ell']  } + \cos{\delta_{cp}} \frac{ w_{cos\ell\ell'}^{ab} }{ {\hat {\bar D}}[\ell]   {\hat {\bar D}}[\ell']  } \nonumber \\
&+& \cos^2{\delta_{cp}} \frac{ w_{cos^2\ell\ell'}^{ab} }{ {\hat {\bar D}}[\ell]   {\hat {\bar D}}[\ell']  } ~.
\eeq 

Equations~(\ref{Fabal1},\ref{a:rate}) show that $F^{ab}_{\ell\ell'}$ plays a role for $\mathcal{P}(\nu_a  \rightarrow \nu_b)$ similar to the one that $F^{ab}_{\ell}$ plays for $S(T)$.
Using Eq.~(\ref{a:Pmabs2}), it follows from Eqs.~(\ref{Pmabs2aI},\ref{Pmabs2aR}) that $Im F^{ab}_{\ell\ell'}$ is odd, and $Re F^{ab}_{\ell\ell'}$ even, under $\delta_{cp}\rightarrow -\delta_{cp}$.

\end{document}